\documentclass[reprint, aip, twoside, twocolumn, floatfix, nofootinbib]{revtex4-1}
\usepackage{amsmath, amssymb}
\usepackage{graphicx}
\usepackage[usenames, dvipsnames]{color}
\usepackage{hyperref}
\hypersetup{colorlinks=true, linkcolor = blue, citecolor = blue, urlcolor = blue}
\newcommand{\dd}{\mathrm{d}}
\newcommand{\ee}{\mathrm{e}}
\newcommand{\ii}{\mathrm{i}}
\newcommand{\Tr}{\mathrm{Tr}}

\newcommand{\cD}{\mathcal{D}}
\newcommand{\cE}{\mathcal{E}}
\newcommand{\cG}{\mathcal{G}}

\newcommand{\cS}{\mathcal{S}}

\newcommand{\Pop}{\mathcal{P}}
\newcommand{\opH}{\hat{\mathcal{H}}}
\newcommand{\oph}{\hat{H}}
\newcommand{\opT}{\hat{\mathcal{T}}}
\newcommand{\opHel}{\hat{\mathcal{H}}_\mathrm{el}}
\newcommand{\opV}{\hat{\mathcal{V}}}
\newcommand{\opv}{\hat{V}}
\newcommand{\opp}{\hat{p}}
\newcommand{\ops}{\hat{\sigma}}

\newcommand{\opA}{\hat{\mathcal{A}}}
\newcommand{\opO}{\hat{\mathcal{O}}}
\newcommand{\Gs}{\mathsf{g}}
\newcommand{\vecp}{\mathbf{p}}
\newcommand{\vecR}{\mathbf{R}}
\newcommand{\vecr}{\mathbf{r}}

\newcommand{\vecF}{\mathbf{F}}
\newcommand{\vecA}{\mathbf{A}}
\newcommand{\avgR}{\bar{\vecR}}
\newcommand{\vecv}{\mathbf{v}}
\newcommand{\avgv}{\bar{\vecv}}

\newcommand{\nmax}{N_\mathrm{nt}}
\everymath{\displaystyle}
\begin{document}
\title{Mixed quantum-classical approach to model non-adiabatic electron-nuclear dynamics: Detailed balance and improved surface hopping method}
\author{E.~V.~Stolyarov}
\affiliation{Institute of Physics of the National Academy of Sciences of Ukraine, pr. Nauky 46, 03028 Kyiv, Ukraine}
 \author{A.~J.~White}
 \email[]{alwhite@lanl.gov}
 \affiliation{Theoretical Division, Los Alamos National Laboratory, Los Alamos, NM 87545, USA}
 \author{D.~Mozyrsky}
 \email[]{mozyrsky@lanl.gov}
 \affiliation{Theoretical Division, Los Alamos National Laboratory, Los Alamos, NM 87545, USA}
\begin{abstract}
 We develop a density matrix formalism to describe coupled electron-nuclear dynamics. To this end we introduce an effective Hamiltonian formalism that describes electronic transitions and small (quantum) nuclear fluctuations along a classical trajectory of the nuclei. Using this Hamiltonian we derive equations of motion for the electronic occupation numbers and for the nuclear coordinates and momenta. We show that in the limit when the number of nuclear degrees of freedom coupled to a given electronic transition is sufficiently high (i.e., the strong decoherence limit), the equations of motion for the electronic occupation numbers become Markovian. Furthermore the transition rates in these (rate) equations are asymmetric with respect to the lower-to-higher energy transitions and vice versa. In thermal equilibrium such asymmetry corresponds to the detailed balance condition. We also study the equations for the electronic occupations in non-Markovian regime and develop a surface hopping algorithm based on our formalism. To treat the decoherence effects we introduce additional ``virtual'' nuclear wavepackets whose interference with the ``real'' (physical) wavepackets leads to the reduction in coupling between the electronic states (i.e., decoherence) as well as to the phase shifts that improve the accuracy of the numerical approach. Remarkably, the same phase shifts lead to the detailed balance condition in the strong decoherence limit.
\end{abstract}
\setcounter{page}{1}
\maketitle
\section{Introduction} \label{sec:intro}
Accurate modeling of coupled electron-nuclear dynamics is key to understanding molecular photophysics and optoelectronics. To model molecules larger than a few atoms, one can use direct nonadiabatic molecular dynamics simulations, where the total molecular energy and forces on the nuclei are calculated \emph{on-the-fly} for all required electronic states \cite{Nelson20}. These  quantities, calculated by time-dependent or linear response electronic structure methods, are used in either a mixed quantum-classical \cite{Tully98} or fully quantum algorithm to calculate the dynamics. The prior often suffer from ``overcoherence" problems due to mean-field treatment of the electronic or electron-nuclear wavefunction\cite{Subotnik16,Barbatti18}. The latter is intractable on fixed high-dimensional grids, necessitating the use of trajectory-guided basis functions\cite{Shalashilin05,Shalashilin08,Habershon12,Habershon15,Grigolo16}. Most often the basis functions are guided through an approximate mean-field \cite{MAKHOV2017200,Shalashilin09,Shalashilin10} or \emph{ad hoc} cloning/spawning algorithm \cite{Makhov14,Yang09}, which does not guarantee completeness of the basis set \cite{Symonds18}. Additionally, for direct dynamics the energy and forces are not known for all nuclear configurations which leads to further approximations in the fully quantum dynamics approach, \emph{i.e.} a mid-point or bra-ket averaged Taylor approximation for matrix elements involving different basis functions \cite{Mignolet18}. Due to the uncertainties in these approaches, there is a need for algorithms which provide the best possible semi-classical dynamics for the trajectories, which also only depend on the energy and forces of the instantaneous nuclear configuration calculated in direct dynamics.

Over the past few years, there have been efforts by ourselves \cite{White16,Baskov2019} and others \cite{Izmaylov18,Pengfei18} to develop non-adiabatic molecular dynamics algorithms which consider the electronic states as strictly defined by the time dependent nuclear configurations. We showed previously that this definition of the electronic states, in the adiabatic representation only, leads directly to the mixed quantum-classical concept of momentum-jumps \cite{White16}, used in surface hopping \cite{Tully1990} and also found in the quantum-classical Liouville equation formalism \cite{Kapral06}. Recently, we refined this concept by defining an effective momentum-jump Hamiltonian and formulated an improved Ehrenfest-like algorithm, which can be employed in cloning/spawning algorithms \cite{Baskov2019}. Here we formulate a density matrix based generalization of the approach in order to 1) propose a tractable, albiet slighlty more complex, surface-hopping-like algorithm to model the coupled electron-nuclear dynamics, which interpolates between the quantum, slightly-decoherent and non-Markovian regime, and the classical, highly-decoherent regime and 2) demonstrate and interpret the realization of detailed balance in the classical regime at finite temperature. The latter is poorly represented by common mixed quantum classical methods \cite{Tully06}, is critical to thermal equilibration, and has been subject of much recent theoretical interest \cite{Mozyrsky18,Sifain16,Subotnik16,Jain15,Kang19,Miller15}.

The structure of the paper is as follows.
In Sec.~\ref{sec:theory} we formulate the effective Hamiltonian to treat the coupled mixed electron-nuclear dynamics. Then we derive the equations of motion for the electronic density matrix as well as equation of motion for the nuclei determined by the effective Hamiltonian. We also show that under the appropriate conditions the equations for the electronic occupations become classical rate (master) equations that obey the property of detailed balance. In Sec.~\ref{sec:algr} we describe the details of algorithm to model non-adiabatic electron-nuclear dynamics based on the equations of motion derived in Sec.~\ref{sec:theory}.
We apply the algorithm to the Tully's test problems and discuss the results in Sec. \ref{sec:num}.
Additional derivations and technical details are delegated to the Appendices.
\section{Theory} \label{sec:theory}
\subsection{The effective Hamiltonian}
The molecular Hamiltonian can be represented as
\begin{equation} \label{eq:full_ham}
 \opH = \opT + \opHel(\vecR).
\end{equation}
The first term corresponds to the kinetic energy of the nuclei $\textstyle \opT=\sum_{\alpha} \opp^2_\alpha/(2M_\alpha)$, with $\textstyle \opp_\alpha = -i \nabla_{R_\alpha}$ being a momentum operator, where $\textstyle R_\alpha$ stands for the $\alpha$-th component of the set of nuclear coordinates $\textstyle \vecR = \left(R_1, R_2, \ldots, R_{3N}\right)$ and $\textstyle N$ is the number of nuclei.
The term $\textstyle \opHel$ stands for the Hamiltonian of the electronic subsystem.
It contains the electronic kinetic energy, all electron-electron, electron-nuclear, and nuclear-nuclear interactions, and parametrically depends on the position of all nuclei $\textstyle \vecR$.
Throughout the paper, we assume atomic units, \emph{i.e.} $\textstyle \hbar = k_{B}=\epsilon_0=1$, for convenience.

The Hamiltonian \eqref{eq:full_ham} is usually represented in the basis of the eigenstates of the electronic Hamiltonian $\textstyle \opHel(\vecR)$:~\cite{Baer2006}
\begin{equation} \label{eq:eigen}
 \opH_\mathrm{el}(\vecR)|n(\vecR)\rangle = E_n(\vecR) |n(\vecR)\rangle,
\end{equation}
where $\textstyle E_n(\vecR)$ is referred to as a potential energy surface (PES) of the adiabatic electronic state $\textstyle |n(\vecR)\rangle$.
Within this basis the molecular Hamiltonian given by Eq.~\eqref{eq:full_ham} transforms into a ``velocity-gauge" Hamiltonian\cite{Baer2006} (see the derivation in Appendix~\ref{sec:hvg_der}).

However, if one works with the adiabatic eigenstates $\textstyle |n(\vecR)\rangle$ which are functions of $\textstyle 3N$-dimensional vector of nuclear coordinates $\textstyle \vecR$, computation of PESs and NACVs quickly becomes prohibitive due to exponential increase in the required computational resources with the increase of the number of the nuclear degrees of freedom.
In these cases, on-the-fly methods for \emph{ab-initio} NAMD simulations are preferable.
In these methods, PESs and NACVs are evaluated at the some time dependent position of nuclei $\textstyle \avgR(t)$.
In this case, one works with time-dependent adiabatic basis states $\textstyle |n(t)\rangle$.
The state of the system $\textstyle |\varPsi(\vecR, t)\rangle$ can be represented as a superposition of these states as
\begin{equation} \label{eq:Psi}
 |\varPsi(\vecR, t)\rangle = \sum_{n} \psi_{n}(\vecR, t) |n(t)\rangle,
\end{equation}
where $\textstyle \psi_n(\vecR, t)$ is a nuclear wavefunction.
Using the expansion~\eqref{eq:Psi} in the time-dependent Schr\"{o}dinger equation
$\textstyle \ii\partial_t|\varPsi(t)\rangle = \opH|\varPsi(t)\rangle$, one obtains
\begin{equation}
 \begin{split}
   & \ii \sum_{n}\left[\partial_t \psi_{n}(\vecR, t)\right]|n(t)\rangle
                        + \ii \sum_{n}\psi_{n}(\vecR, t)|\partial_t n(t)\rangle \\
   & \, = \sum_{n'} \opH \psi_{n}(\vecR, t)|n(t)\rangle.
 \end{split}
\end{equation}
\begin{equation}
 \ii \partial_t \psi_{n'}(\vecR, t) = \sum_{n} h_{n'n}\psi_{n}(\vecR, t),
\end{equation}
where $\textstyle h_{n'n} = \langle n'(t)|\opH|n(t)\rangle - \ii \langle n'(t)|\partial_t n(t)\rangle$.
This leads to the representation of the molecular Hamiltonian in a ``length gauge" \cite{Baskov2019}
\begin{equation} \label{eq:ham_vg}
 \opH_\mathrm{lg} = \opT +
 \sum_{n,n'} \left[\ii \langle n(t)|\partial_t n'(t)\rangle + U_{nn'}(t)\right]|n'(t)\rangle\langle n(t)|,
\end{equation}
where $\textstyle U_{nn'}(t) = \langle n(t)|\opH_\mathrm{el}(\vecR)|n'(t)\rangle$.
Here we assume that no magnetic field is applied to the system.
In this case, the electronic Hamiltonian $\textstyle \opHel$ is real and $\textstyle U_{nn'}(t)=U_{n'n}(t)$.

To proceed, we speculate that the nuclear wavefunctions, $\textstyle \psi_n(\vecR,t)$, are strongly localized in the vicinity of $\textstyle \avgR(t)$.
We expand the electronic Hamiltonian $\textstyle \opHel$ around $\textstyle \avgR$ retaining only the zero- and the first-order terms
\begin{equation} \label{eq:Hel_taylor}
 \opHel(\vecR) \approx \opHel(\avgR) + \nabla_{\avgR} \opHel(\avgR) \cdot(\vecR - \avgR).
\end{equation}
The higher-order terms are assumed to be negligible for a sufficiently localized nuclear wavefunction.

Using Eqs.~\eqref{eq:Hel_taylor} and \eqref{eq:eigen} along with the Hellmann-Feynman theorem~\cite{Levine2014}, 
one obtains for $\textstyle U_{nn'}$ the result as follows
\begin{equation} \label{eq:unn}
 \begin{split}
    U_{nn'} = & \, \delta_{nn'}\left[E_n(\avgR) - \vecF_n(\avgR)\cdot(\vecR-\avgR)\right] \\
    & \, + (1-\delta_{nn'}) \Delta E_{nn'}(\avgR) \vecA_{nn'}\cdot(\vecR-\avgR),
 \end{split}
\end{equation}
where $\textstyle \vecA_{nn'}(\vecR) \equiv \langle n(\vecR)|\nabla_\vecR n(\vecR)\rangle$ is referred to as a non-adiabatic coupling vector (NACV) and $\textstyle \vecF_n(\vecr) = - \nabla_\vecr E_n(\vecr)$ is the force acting on a nucleus on the $n$-th PES at coordinate $\textstyle \vecr$.
\begin{equation} \label{eq:ndtn}
 \langle n(\avgR)|\partial_t n'(\avgR)\rangle
 = \dot{\avgR}\cdot\langle n(\avgR)|\nabla_{\avgR}  n'(\avgR)\rangle
 = \avgv \cdot \vecA_{nn'}(\avgR),
\end{equation}
where $\textstyle \avgv = \dot{\avgR}$ is the velocity of the center of the nuclear wavepacket.

Substituting Eqs.~\eqref{eq:unn} and \eqref{eq:ndtn} into the Hamiltonian~\eqref{eq:ham_lg}, one obtains $\textstyle \opH_\mathrm{lg} = \opH_0 + \opV$,
where $\textstyle \opH_0$ is the diagonal part given by
\begin{equation}
 \opH_0 = \opT + \sum_{n} \left[E_n(\avgR) - \vecF_n(\avgR)\cdot(\vecR-\avgR)\right]|n\rangle\langle n|.
\end{equation}
The off-diagonal part $\textstyle \opV$ reads
\begin{gather} \label{eq:lg_offdiag}
 \begin{split}
   \opV =  & \, \sum_{n,n'} \left[\ii \avgv\cdot\vecA_{nn'}(\avgR) \right. \\
   & \, \left. + \Delta E_{nn'}(\avgR) \vecA_{nn'}(\avgR) \cdot (\vecR - \avgR)\right]|n'\rangle\langle n| \\
   \equiv  & \, \ii \sum_{n,n'} \avgv\cdot\vecA_{nn'}(\avgR)\left[1 + \ii \Delta \vecp_{nn'} \cdot (\vecR - \avgR) \right]|n'\rangle\langle n|,
 \end{split}
\end{gather}
where $\textstyle \Delta E_{nn'}(\bar{\bf R}) = E_{n}(\bar{\bf R}) - E_{n'}(\bar{\bf R})$ is the energy difference between $n$-th and $n'$-th PESs at coordinate $\bar{\bf R}$ and
\begin{equation} \label{eq:Delta_p}
\Delta \vecp_{nn'} = - \Delta E_{nn'} \vecA_{nn'}/(\vecv \cdot \vecA_{nn'}).
\end{equation}
Hereinafter, for brevity, we use a shorthand notation $\textstyle |n\rangle = |n(t)\rangle$.

Since the nuclear wavefunction is strongly localized near $\textstyle \avgR$, the condition $\textstyle \Delta \vecp_{nn'}\cdot(\vecR - \avgR) \ll 1$ holds.
In this case, the last line in Eq.~\eqref{eq:lg_offdiag} can be approximated as
\begin{equation} \label{eq:lg_offdiag_exp}
 \opV \approx \ii \sum_{n,n'} \avgv \cdot \vecA_{nn'}(\avgR) \ee^{\ii \Delta\vecp_{nn'}\cdot(\vecR-\avgR)} |n'\rangle\langle n|.
\end{equation}

The Hamiltonian in Eq.~\eqref{eq:lg_offdiag_exp} is a much more efficient representation of the interaction between the electronic states and the nuclei than the interaction Hamiltonian in Eq.~\eqref{eq:lg_offdiag} for both physical and computational reasons.  If we assume the nuclear wavefunction takes a Gaussian form, a very common semiclassical anzatz that we will be using below, the application of the Hamiltonian~\eqref{eq:lg_offdiag_exp} to the nuclear state leaves it in Gaussian form, albiet with a new momentum and amplitude. Application of the Hamitonian in Eq.~\eqref{eq:lg_offdiag} transforms a Gaussian wavefunction into the first order Hermite polynomial. If the Hamiltonian in Eq.~\eqref{eq:lg_offdiag_exp} is applied twice, the Gaussian wavepacket remains the same (of course, except for its amplitude), while the dual application of the Hamiltonian~\eqref{eq:lg_offdiag_exp} leads to the superposition of first and second order Hermite polynomials. The latter wavefunction does not have a straightforward classical analog, unlike a Gaussian wavepacket which corresponds to the semiclassical particle strongly confined around its average (classical) position and momentum. Thus the interaction Hamiltonian in Eq.~\eqref{eq:lg_offdiag_exp} tends to maintain the ``classicality'' of the nuclei, which is important both for physical interpretation and numerical modeling. Indeed, application of Hamiltonian in Eq.~\eqref{eq:lg_offdiag} requires extended basis set of Hermite polynomials that grows exponentially with the number of the nuclear degrees of freedom.

Using the approximation~\eqref{eq:lg_offdiag_exp}, we finally arrive at the \emph{momentum-jump Hamiltonian} \cite{Baskov2019}
\begin{equation} \label{eq:ham_mj}
 \begin{split}
   \opH_\mathrm{mj} = & \, \opT + \sum_n \left[E_n(\avgR) - \vecF(\avgR)\cdot(\vecR-\avgR)\right]|n\rangle\langle n| \\
   & \, + \ii \sum_{n,n'} \avgv \cdot \vecA_{nn'}(\avgR) \ee^{\ii \Delta\vecp_{nn'}\cdot(\vecR-\avgR)}|n'\rangle\langle n|.
 \end{split}
\end{equation}
This is the Hamiltonian we work with through the rest of the paper.
\subsection{Evolution equations for adiabatic states populations}
In this subsection we derive equations of motion for the electronic occupation numbers. For simplicity we will assume that we are dealing with a level crossing of two electronic PES. Such situation is quite generic for molecular systems, and the extension to larger number of the PESs is rather straightforward if deemed necessary.  

The density matrix $\textstyle \hat{\rho}(t) = |\Psi(t)\rangle\langle \Psi(t)|$ describing the nuclear state is governed by the von Neumann equation
\begin{equation}
 \ii \partial_t \hat{\rho} = \left[\opH_\mathrm{mj}, \hat{\rho}\right].
\end{equation}
Let us proceed to the interaction picture
\begin{equation} \label{eq:int_pict}
 \hat{\varrho}(t) = \ee^{\ii \opH_0 t} \hat{\rho}(t) \ee^{-\ii \opH_0 t}.
\end{equation}
The evolution equations for the diagonal elements of the density matrix $\textstyle \varrho_{nn}$ are given by
\begin{subequations}
 \begin{equation} \label{eq:eq_r11}
  \ii \partial_t \varrho_{11} = \ee^{\ii \oph_1 t} \opv_{12} \ee^{-\ii \oph_2 t} \varrho_{21}
    - \varrho_{12} \ee^{\ii \oph_2 t} \opv_{21} \ee^{-\ii \oph_1 t},
 \end{equation}
 \begin{equation} \label{eq:eq_r22}
  \ii \partial_t \varrho_{22} = \ee^{\ii \oph_2 t} \opv_{21} \ee^{-\ii \oph_1 t} \varrho_{12}
   - \varrho_{21} \ee^{\ii \oph_1 t} \opv_{12} \ee^{-\ii \oph_2 t},
 \end{equation}
\end{subequations}
where we have introduced the following notations
\begin{subequations} \label{eq:def_hv}
  \begin{gather}
   \oph_n \equiv \left[\opT +E_n(\avgR) - \vecF_n(\avgR)\cdot(\vecR-\avgR)\right] \ops_{nn}, \\
   \opv_{nn'} \equiv \ii \avgv \cdot \vecA_{nn'}(\avgR) \ee^{\ii \Delta\vecp_{nn'}\cdot(\vecR-\avgR)} \ops_{nn'},
  \end{gather}
\end{subequations}
with $\ops_{nn'} \equiv |n\rangle\langle n'|$ being a ladder operator.

The non-diagonal elements of the density matrix (coherences) are governed by
\begin{equation} \label{eq:eq_r12}
  \ii \partial_t \varrho_{12} = \ee^{\ii \oph_1 t} \opv_{12} \varrho_{22} \ee^{-\ii \oph_2 t} - \ee^{\ii \oph_1 t} \varrho_{11} \opv_{12} \ee^{-\ii \oph_2 t}.
\end{equation}
The evolution equation for $\textstyle \varrho_{21}$ is derived by the Hermitian conjugation of the above expression.

The populations of the adiabatic states are expressed as $\textstyle \Pop_n(t) = \Tr[\rho_{nn}(t)] = \Tr[\varrho_{nn}(t)]$, where the trace is taken over all nuclear degrees of freedom.
Substituting the formal solution of Eq.~\eqref{eq:eq_r12} into Eq.~\eqref{eq:eq_r11} assuming that $\textstyle \varrho_{12}(0) = \varrho_{21}(0) = 0$ and taking the trace, one obtains
\begin{widetext}
 \begin{equation} \label{eq:eq_P1}
   \dot{\Pop}_1 = - 2\mathrm{Re}\Big\{\int^t_0 \dd\tau \, \mathrm{Tr}
  \big[\ee^{-\ii H_1 (t-\tau)} V_{12}(t) \ee^{-\ii H_2 (t-\tau)} \opv_{21}(\tau)\rho_{11}(\tau)
       - \opv_{21}(\tau)\ee^{\ii H_1 (t-\tau)} \opv_{12}(\tau) \ee^{-\ii H_2(t-\tau)}\rho_{22}(\tau)\big]\Big\},
 \end{equation}	
where we used Eq.~\eqref{eq:int_pict} and the cyclic permutation property of the trace.
The equation of motion for $\textstyle \Pop_2(t)$ is derived in the analogous manner.

Using that $\textstyle \Tr\left[\opO \rho_{nn}(t)\right] = \langle \Psi_n(t)|\opO|\Psi_n(t)\rangle$,
where $\textstyle \opO$ stands for a quantum-mechanical operator, Eq.~\eqref{eq:eq_P1} takes the form
\begin{equation} \label{eq:eq2_P1}
 \begin{split}
   \dot{\Pop}_1 = & \, 2 d_{12}(t) \int^t_0 \dd\tau \,  d_{12}(\tau) \\
   & \, \times \mathrm{Re}
   \left[\langle \Psi_1(\tau)|\ee^{\ii H_1 (t-\tau)} \ops_{12}(t) \ee^{-\ii H_2 (t-\tau)} \ops_{21}(\tau)|\Psi_1(\tau)\rangle
   - \langle \Psi_2(\tau)|\ops_{21}(\tau)\ee^{\ii H_1 (t-\tau)} \ops_{12}(t) \ee^{-\ii H_2(t-\tau)}|\Psi_2(\tau)\rangle\right],
 \end{split}
\end{equation}
where we have introduced a notation $\textstyle d_{12}(t) \equiv \bar{v}(t) A_{12}(\bar{R}(t))$ and used Eq.~\eqref{eq:def_hv} and the property $\textstyle A_{nn'} = - A_{n'n}$.

Let us invoke here the Gaussian anzats by assuming the nuclear state to be of the form
\begin{equation} \label{eq:GaussA}
 |\Psi_n(t)\rangle = C_n(t) |\Gs_n(\bar{R},\bar{p}; R, t)\rangle,
\end{equation}
where
\begin{equation} \label{eq:GaussS}
 \begin{split}
  & |\Gs_n(\bar{R},\bar{p}; R, t)\rangle = \cG_n(\bar{R},\bar{p}; R, t) |n(t)\rangle, \\
  & \langle \Gs_n(\bar{R},\bar{p}; R, t)|\Gs_n(\bar{R},\bar{p}; R, t)\rangle = 1
 \end{split}
\end{equation}
stands for the normalized Gaussian state with
 \begin{equation} \label{eq:GaussW}
  \cG_n(\bar{R}, \bar{p}; R, t) = \exp\left[\ii \alpha_n(t)\left[R -\bar{R}(t)\right]^2 + \ii \bar{p}(t)\left[R-\bar{R}(t)\right] + \ii \gamma_n(t)\right].
 \end{equation}
\end{widetext}
The Gaussian wave packet $\textstyle \cG_n(\bar{R}, \bar{p}; R, t)$ is localized around the position $\textstyle \bar{R}(t)$ and propagates along the $n$-th PES with the average momentum $\bar{p}(t)$.
The average momentum $\bar{p}(t)$ and position $\bar{R}(t)$ of the wave packet are governed by the classical equations of motion.

In this subsection, to simplify notation, we restrict ourselves to the one-dimensional case. Extension to the multidimensional case (i.e. to multi-atomic systems) is straightforward, particularly within the approximations utilized in this work. We will carry it out when needed, particularly in the next subsection.

Complex parameters $\alpha_n(t)$ and $\gamma_n(t)$ evolve according to the Heller's equations of motion \cite{Heller1975}
\begin{subequations}
 \begin{equation} \label{eq:eq_alpha}
  \dot{\alpha}_n(t) = -\frac{2}{M} \alpha^2(t) - \frac{1}{2} \frac{\partial^2 E_n(R)}{\partial R^2}\Big|_{R=\bar{R}}.
 \end{equation}
 \begin{equation} \label{eq:eq_gamma}
  \dot{\gamma}_n(t) = \frac{\ii}{M} \alpha_n(t) + \frac{\bar{p}^2(t)}{2M} - E_n(\bar{R}(t)).
 \end{equation}
\end{subequations}
Next, following the assumption of strong localization of the wave packet pivotal in the derivation of the momentum-jump Hamiltonian $\textstyle \opH_\mathrm{mj}$, we drop the second term on the right-hand side (rhs) of Eq.~\eqref{eq:eq_alpha}.
The solution of this equation reads
\begin{equation} \label{eq:alpha_sln}
 \alpha_n(t) \equiv \alpha(t) = \frac{\alpha_0}{2 \alpha_0 M^{-1} t + 1}, \quad \alpha_0 = \frac{\ii}{2\sigma^2},
\end{equation}
which corresponds to the case of a free wave packet propagation.
The solution of Eq.~\eqref{eq:eq_gamma} reads as~\cite{Heller1975}
\begin{equation} \label{eq:gamma_sln}
 \gamma_n(t) = \gamma_0 + \frac{\ii}{2} \ln\left(\frac{2}{M} \alpha_0 t + 1\right) + \cS_n(t,0),
\end{equation}
where $\gamma_0 = \ii \ln\left(2\alpha_0/\pi\right)/4$ and
\begin{equation} \label{eq:def_actn}
  \cS_n(t,0) = \int^t_0 \dd u \left[\frac{\bar{p}^2(u)}{2M} - E_n(\bar{R}(u))\right]
\end{equation}
stands for a classical action produced by a wave packet during time $t$.

\begin{figure*}[t!]
	\centering
	\includegraphics{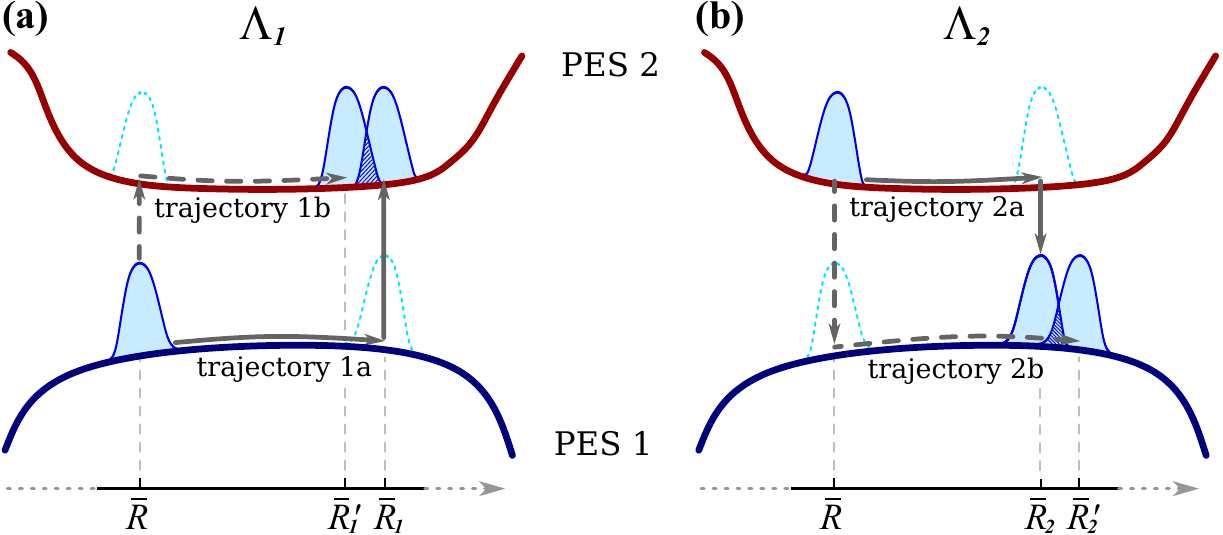}
	\caption{Graphical representation of the trajectories contributing to correlators (a) $\Lambda_1(t;\tau)$ and (b) $\Lambda_2(t;\tau)$.
		\label{fig:fig_paths}}
\end{figure*}

Using the anzats given by Eq.~\eqref{eq:GaussA} in Eq.~\eqref{eq:eq2_P1} and taking into account that $\textstyle \Pop_n(t) = |C_n(t)|^2$, one arrives at a set of equations of motion for the adiabatic states populations as follows
\begin{subequations} \label{eq:pops_eqs}
 \begin{equation} \label{eq:p1_eq1}
  \dot{\Pop}_1(t) = \int^t_0 \dd\tau \mathrm{Re}\left[Q_2(t;\tau) \Pop_2(\tau) - Q_1(t;\tau) \Pop_1(\tau)\right],
 \end{equation}
 \begin{equation} \label{eq:p2_eq1}
  \dot{\Pop}_2(t) = \int^t_0 \dd\tau \mathrm{Re}\left[Q_1(t;\tau) \Pop_1(\tau) - Q_2(t;\tau) \Pop_2(\tau)\right],
 \end{equation}
\end{subequations}
where $\textstyle Q_1(t;\tau)$ and $\textstyle Q_2(t;\tau)$ are determined as
\begin{subequations}
	\begin{equation}
	Q_1(t;\tau) = 2 d_{12}(t) d_{12}(\tau) \Lambda_1(t;\tau),
	\end{equation}
	\begin{equation}
	Q_2(t;\tau) = 2 d_{12}(t) d_{12}(\tau) \Lambda_2(t;\tau),
	\end{equation}
\end{subequations}
with $\textstyle \Lambda_{1,2}(t;\tau)$ given by
\begin{widetext}
\begin{subequations} \label{eq:def_Q}
  \begin{equation} \label{eq:def_Q12}
  \Lambda_1(t;\tau) = \langle \Gs_1(\bar{R},\bar{p};R,\tau)|\overbrace{\ee^{\ii \oph_1 (t-\tau)} \ops_{12} (t)}^{\text{trajectory 1a}}
  \overbrace{\ee^{-\ii \oph_2 (t-\tau)} \ops_{21}(\tau)}^{\text{trajectory 1b}}|\Gs_1(\bar{R},\bar{p};R,\tau)\rangle,
  \end{equation}
  \begin{equation} \label{eq:def_Q21}
   \Lambda_2(t;\tau) = \langle \Gs_2(\bar{R},\bar{p};R,\tau)|\overbrace{\ops_{21}(\tau) \ee^{\ii \oph_1 (t-\tau)}}^{\text{trajectory 2b}}
   \overbrace{\ops_{12}(t) \ee^{-\ii \oph_2 (t-\tau)}}^{\text{trajectory 2a}}|\Gs_2(\bar{R},\bar{p};R,\tau)\rangle.
  \end{equation}
\end{subequations}
\end{widetext}
Using Eqs.~\eqref{eq:GaussS} and \eqref{eq:GaussW}, Eqs.~\eqref{eq:def_Q} transform into
\begin{subequations} \label{eq:overlaps}
 \begin{equation}
  \begin{split}
   \Lambda_1(t;\tau) = \langle \Gs_2(\bar{R}_1, \bar{p}_1; R, t)|\Gs_2(\bar{R}'_1, \bar{p}'_1; R, t)\rangle,
  \end{split}
 \end{equation}
 \begin{equation}
   \Lambda_2(t;\tau) = \langle \Gs_1(\bar{R}_2, \bar{p}_2; R, t)|\Gs_1(\bar{R}'_2, \bar{p}'_2; R, t)\rangle,
 \end{equation}
\end{subequations}
which correspond to the overlaps of pairs of Gaussian wave packets.

In the case of $\Lambda_1(t;\tau)$, these wave packets are results of propagation of the Gaussian wave packet (with momentum $\bar{p}$ and position $\bar{R}$) residing on PES 1 at time $\tau$ over two different trajectories (1a and 1b). The schematic representation of these trajectories is demonstrated in Fig.~\ref{fig:fig_paths}a.
Trajectory 1a corresponds to propagation of the wave packet along PES 1 from position $\bar{R}(\tau)$ to $\bar{R}_1(t)$ and a jump to PES 2 at an instant $t$. That is, at the end of this trajectory the wave packet  momentum changes by $\Delta  p_{12}(t)$.
Trajectory 1b consists of a jump from PES 1 to PES 2 at an instant $\tau$  (accompanied by an instanteneous change in momentum by $\Delta p_{12}(\tau)$) and propagation over PES 2 from position $\bar{R}(\tau)$ to $\bar{R}'_1(t)$ during the interval of time $[\tau, t]$.

Similarly, for $\Lambda_2(t;\tau)$, we deal with the overlap of wave packets obtained as a result of propagation of the Gaussian wave packet residing on PES 2 at an instant $\tau$ over trajectories 2a and 2b (see Fig.~\ref{fig:fig_paths}b).
Trajectory 2a consists of a piece corresponding to propagation over PES 2 during the interval $[\tau, t]$ from $\bar{R}(\tau)$ to $\bar{R}_2(t)$ followed by a jump to PES 1 at an instant $t$.
Again the momentum of the wavepacket changes by $\Delta p_{21}(t)$  at the end of trajectory 2a.
Trajectory 2b corresponds to a jump from PES 2 to PES 1 at $\tau$ and propagation over PES 1 from $\bar{R}(\tau)$ to $\bar{R}'_2(t)$.

\begin{widetext}
Using Eq.~\eqref{eq:GaussW} along with Eqs.~\eqref{eq:gamma_sln} and \eqref{eq:def_actn} in Eq.~\eqref{eq:overlaps}, one obtains
\begin{subequations} \label{eq:L_slns}
 \begin{equation}
  \Lambda_1(t;\tau) = \exp\left[- \ii \varphi_{12}(t;\tau)\right] \exp\left[-\frac{1}{4\sigma^2}\Big(\bar{R}_1(t) - \bar{R}'_1(t)-(t-\tau)\frac{(\bar{P}_1(t) - \bar{P}'_1(t))}{M}\Big)^2 -\frac{\sigma^2}{4}\left(\bar{P}_1(t) - \bar{P}'_1(t)\right)^2\right],
 \end{equation}
 \begin{equation}
  \Lambda_2(t;\tau) = \exp\left[- \ii \varphi_{21}(t;\tau)\right] \exp\left[-\frac{1}{4\sigma^2}\Big(\bar{R}_2(t) - \bar{R}'_2(t)-(t-\tau)\frac{(\bar{P}_1(t) - \bar{P}'_1(t))}{M}\Big)^2 -\frac{\sigma^2}{4}\left(\bar{P}_2(t) - \bar{P}'_2(t)\right)^2\right],
 \end{equation}
\end{subequations}
where $\bar{P}_1(t)$, $\bar{P}'_1(t)$, $\bar{P}_2(t)$ and $\bar{P}'_2(t)$  are the final momenta of trajectories 1a, 1b, 2a and 2b, respectively.

The phases $\textstyle \varphi_{12}(t;\tau)$ and $\textstyle \varphi_{21}(t;\tau)$ in Eqs.~\eqref{eq:L_slns} are given by
\begin{subequations} \label{eq:phases}
	\begin{equation}
		\varphi_{12}(t;\tau) = \int^t_\tau \dd u \left[\frac{\bar{p}_\mathrm{1a}^2(u)}{2M} - E_1(\bar{R}_\mathrm{1a}(u))\right]
		- \int^t_\tau \dd u \left[\frac{\bar{p}_\mathrm{1b}^2(u)}{2M} - E_2(\bar{R}_\mathrm{1b}(u))\right]
		- \frac{1}{2}\left[\bar{P}_1(t) + \bar{P}'_1(t)\right]\left[\bar{R}_1(t)-\bar{R}'_1(t)\right],
	\end{equation}
	\begin{equation}
		\varphi_{21}(t;\tau) = \int^t_\tau \dd u \left[\frac{\bar{p}_\mathrm{2a}^2(u)}{2M} - E_2(\bar{R}_\mathrm{2a}(u))\right]
		- \int^t_\tau \dd u \left[\frac{\bar{p}_\mathrm{2b}^2b(u)}{2M} - E_1(\bar{R}_\mathrm{2b}(u))\right]
		- \frac{1}{2}\left[\bar{P}_2(t) + \bar{P}'_2(t)\right]\left[\bar{R}_2(t)-\bar{R}'_2(t)\right],
	\end{equation}
\end{subequations}
\end{widetext}

Several comments should be made regarding Eqs.~\eqref{eq:phases}. The average wavepacket  positions and momenta propagate according to the classical equations
of motion according to the Hamiltonian $\oph_n$  in Eq.~\eqref{eq:def_hv}. This Hamiltonian depends parametrically on $\bar{R}_n$, which is the reference point around which we expand the electronic Hamiltonian $\opHel(\vecR)$ in order to derive  Eq.~\eqref{eq:ham_mj}. Note that the states $|\Gs_1(\bar{R}_1, \bar{p}_1; R, t)\rangle$ and
$|\Gs_2(\bar{R}_2, \bar{p}_2; R, t)\rangle$ are different, hence $\bar{R}_1$ and  $\bar{R}_2$  are different. While their choice, generally speaking, is arbitrary (as long as the expansion in Eq~\eqref{eq:Hel_taylor} is accurate enough), it is reasonable to choose that  $\bar{R}_1$ coincides with either $\bar{R}_\mathrm{1a}$ or $\bar{R}_\mathrm{1b}$ (here we choose $\bar{R}_1=\bar{R}_\mathrm{1a}$), and similarly we set $\bar{R}_2=\bar{R}_\mathrm{2a}$. One can then argue that the potential energies $E_2$ for the trajectories 1b and 2b must be evaluated at points $\bar{R}_\mathrm{1a}(u)$ and $\bar{R}_\mathrm{2a}(u)$, not $\bar{R}_\mathrm{1b}(u)$ and $\bar{R}_\mathrm{2b}(u)$. However, this is not so: Effective Hamiltonians, e.g. Eq.~\eqref{eq:def_hv}, associated with these trajectories contain extra $c$-number time-dependent terms $F_2(\bar{R}_\mathrm{1a}(u))(\bar{R}_\mathrm{1a}(u)-\bar{R}_\mathrm{1b}(u))$ and $F_2(\bar{R}_\mathrm{2a}(u))(\bar{R}_\mathrm{2a}(u)-\bar{R}_\mathrm{2b}(u))$, which arise due to the fact that the energies and forces are evaluated at points $\bar{R}_\mathrm{1a}(u)$ and $\bar{R}_\mathrm{2a}(u)$, respectively. These terms combine with $E_2(\bar{R}_\mathrm{1a}(u))$ and $E_2(\bar{R}_\mathrm{2a}(u))$ shifting the latter to $E_2(\bar{R}_\mathrm{1b}(u))$ and $E_2(\bar{R}_\mathrm{2b}(u))$, respectively, which is accounted for in the Eqs.~\eqref{eq:phases}. Furthermore,
using that
\begin{subequations}\label{coordinates}
\begin{equation}
\bar{R}_\mathrm{1a}(\tau) = \bar{R}_\mathrm{1b}(\tau),\ \bar{R}_\mathrm{1a}(t) = \bar{R}_1(t),\ \bar{R}_\mathrm{1b}(t)= \bar{R}'_1(t),
\end{equation}
\begin{equation}
\bar{p}_\mathrm{1a}(t)+\Delta p_{12}(t) = \bar{P}_1(t),\ \text{and} \,  \bar{p}_\mathrm{1b}(t) = \bar{P}'_1(t),
\end{equation}
\end{subequations}
(as well as analogous relations for the coordinates and momenta for the 2a and 2b trajectories), we integrate the kinetic terms by parts. The boundary term partially cancels the last term (in both expressions) in Eq.~\eqref{eq:phases}, leaving a term
\begin{equation}\nonumber
\Delta p_{12}(\bar{R}_1(t)) \left[\bar{R}_1(t)-\bar{R}'_1(t)\right]/2
\end{equation}
in the rhs of the expression for $\varphi_{12}(t;\tau)$ in Eq.~\eqref{eq:phases} and an analogous expression for $\varphi_{21}(t;\tau)$, but with indices $1$ and $2$ interchanged. For the remaining integral (in the expression for $\varphi_{12}(t;\tau)$) we have
\begin{equation}\nonumber
\begin{split}
&\int^t_\tau \dd u \big(\dot{\bar{p}}_\mathrm{1a}(u)+\dot{\bar{p}}_\mathrm{1b}(u)\big)\,
\frac{\big(\bar{R}_\mathrm{1a}(u)-\bar{R}_\mathrm{1b}(u)\big)}{2}\\
&= \int^t_\tau \dd u \big(F_1(\bar{R}_\mathrm{1a}(u))  +  F_2(\bar{R}_\mathrm{2a}(u))\big)\,
\frac{\big(\bar{R}_\mathrm{1a}(u)-\bar{R}_\mathrm{1b}(u)\big)}{2}\,,
\end{split}
\end{equation}
where we have utilized the fact that the average coordinates and momenta evolve according to
the Newtonian equations of motion. This last expression can be combined with the integrals over the potential
energies in Eq.~\eqref{eq:phases}, and so we obtain
\begin{subequations} \label{eq:phases_opt}
\begin{equation}
\varphi_{12}(t;\tau) = \int^t_\tau \dd u\, \Delta E_{12}(R_{\mathrm{c}1}(u)) - \frac{\Delta p_{12}(\bar{R}_1(t)) \Delta R_1(t)}{2}\,,
\end{equation}
\begin{equation}
\varphi_{21}(t;\tau) = \int^t_\tau \dd u\, \Delta E_{21}(R_{\mathrm{c}2}(u)) - \frac{\Delta p_{21}(\bar{R}_2(t)) \Delta R_2(t)}{2}\,,
\end{equation}
\end{subequations}
where $\Delta R_n(t)=\bar{R}_n(t)-\bar{R}'_n(t)$, $R_{\mathrm{c}n} = (\bar{R}_{n\mathrm{a}} + \bar{R}_{n\mathrm{b}})/2$, and $\Delta E_{nn'}(R)=E_n(R)-E_{n'}(R)$ with $n, n' \in \{1,2\}$.

For the momentum differences at the end of trajectories, we set
\begin{equation} \label{eq:fin_p}
\bar{P}_n(t) - \bar{P}'_n(t) \approx 0, \quad n \in \{1,2\}.
\end{equation}
This relation is exact, of course, only for flat PESs. However, for non-flat PES we note that the difference in the impulses of forces on different PESs,
\begin{equation}\nonumber
\int^t_\tau \dd u \big(F_1(\bar{R}_\mathrm{1a}(u))  -  F_2(\bar{R}_\mathrm{2a}(u))\big)\,,
\end{equation}
is compensated, at least partially, by the difference in $\Delta p_{12}(t)$ and $\Delta p_{12}(\tau)$, e.g. Eq.~\eqref{eq:Delta_p}.

Then we arrive at the expression
\begin{equation} \label{eq:L_slns12}
\Lambda_{12}(t, \tau) \approx  \exp\left\{- \frac{\Delta R_1^2(t)}{4\sigma^2}\right\}\exp\left[\ii \phi_{12}(t;\tau)\right]\,,
\end{equation}
where $\phi_{12}$ is given by Eq.~\eqref{eq:phases_opt}.

The expression for $\Lambda_{21}$ is analogous
\begin{equation} \label{eq:L_slns21}
\Lambda_{21}(t, \tau) \approx  \exp\left\{- \frac{\Delta R_2^2(t)}{4\sigma^2}\right\}\exp\left[\ii \phi_{21}(t;\tau)\right]\,.
\end{equation}

In the following we set $R_{\mathrm{c}2}=R_{\mathrm{c}1}\equiv R_\mathrm{c}$ and $\Delta R_2 = -\Delta R_1\equiv \Delta R$. Indeed, in the semiclassical limit and for a relatively small region of non-adiabatic coupling 
 we expect that the trajectories 1b and 2a (as well as 1a and 2b), see Fig.~\ref{fig:fig_paths}, to be very close to each other (except for the and points at $\tau$ and $t$, of course). This is a consequence of Frank-Condon principle \cite{Landau}, that states that in the vicinity of level crossing the coordinates and momenta of the nuclear states $\Psi_1(\tau)$ and $\Psi_2(\tau)$ (or $|\Gs_1\rangle$ and $|\Gs_2\rangle$ wavepackets) must be the same. Then we can write
 
\begin{widetext}

\begin{subequations}
 \begin{equation} \label{eq:p1_eq2}
    \dot{\Pop}_1(t) = - 2 d_{12}(t) \int^t_0 \dd\tau \, d_{12}(\tau) \cD(t;\tau)
    \left\{\cos\left[\varphi_{12}(t;\tau)\right]\Pop_1(\tau)
    - \cos\left[\varphi_{21}(t;\tau)\right]\Pop_2(\tau)\right\}\,,
 \end{equation}
 \begin{equation} \label{eq:p2_eq2}
    \dot{\Pop}_2(t) = - 2 d_{12}(t) \int^t_0 \dd\tau \, \cD(t;\tau)
     \left\{\cos\left[\varphi_{21}(t;\tau)\right]\Pop_2(\tau)
     - \cos\left[\varphi_{12}(t;\tau)\right]\Pop_1(\tau)\right\}\,,
 \end{equation}
\end{subequations}
with
\begin{equation}\label{eq:D}
\cD(t;\tau) = \exp\left\{- \frac{\big[ R_{\mathrm{r}}(t)-R_{\mathrm{v}}(t,\tau)\big]^2}{4\sigma^2}\right\}\,,
\end{equation}
\begin{equation}\label{eq:phifin}
\varphi_{nn'}(t;\tau) = \int^t_\tau \dd u\, \Delta E_{nn'}(R_\mathrm{c}(u)) - \frac{\Delta p_{12}(R_\mathrm{c}(t))
\big[R_{\mathrm{r}}(t)-R_{\mathrm{v}}(t,\tau)\big]}{2}\,,
\end{equation}
\end{widetext}
and $R_\mathrm{c}(t)= (R_{\mathrm{r}}(t)+R_{\mathrm{v}}(t,\tau))/2$, where, from now on, the subscripts $\mathrm{r}$ and $\mathrm{v}$ label ``real'' 1a (2b) and ``virtual'' 1b (2a) trajectories, respectively. Note that in Eq.~\eqref{eq:phifin} we have set the argument of $\Delta p_{12}$ to the midpoint $R_\mathrm{c}$, which is within the accuracy of the effective Hamiltonian in Eq.~\eqref{eq:ham_mj}. The gaussian $\cD(t;\tau)$ in Eqs.~\eqref{eq:p1_eq2}~\eqref{eq:p2_eq2}  describes the effective reduction of the non-adiabatic coupling due to the reduced wavepacket overlap.

We emphasize that $R_{\mathrm{v}}(t)$ consists of two ``pieces'': $R_{\mathrm{v}}(0,\tau)$, which is a part of trajectory where the virtual wavepacket coincides with the real one ($R_{\mathrm{v}}(0,\tau)= R_{\mathrm{r}}(\tau)$, for $\tau < t$), and $R_{\mathrm{v}}(\tau, t)$, where after having hopped on another PES, the wavepacket propagates on that new PES until time $t$, $R_{\mathrm{v}}(t) = R_{\mathrm{r}}(\tau) + R_{\mathrm{v}}(\tau, t)$. In general, a rigorous evaluation of $R_{\mathrm{v}}(t,\tau))$, though possible, yet, represents a serious computational expense.
In the limit of strong decoherence, when the first exponent in Eq.~\eqref{eq:L_slns12} is very rapidly decaying for $\tau\neq t$, we can use an estimate
\begin{equation} \label{eq:fin_DeltaR1}
R_{\mathrm{r}}(t)-R_{\mathrm{v}}(t,\tau) \approx \frac{\Delta p_{12}(R_\mathrm{c}(t))}{M}\, (t-\tau)\,.
\end{equation}
This estimate is used in the next subsection, where we will derive rate equations using expressions in Eqs.~\eqref{eq:L_slns12} and~\eqref{eq:phases_opt}. For quantum coherent dynamics we use another simple prescription for evaluating the difference:
\begin{equation} \label{eq:fin_DeltaR2}
R_{\mathrm{r}}(t)-R_{\mathrm{v}}(t,\tau) \approx R_{\mathrm{r}}(t)-R_{\mathrm{v}}(t,0)  \,,
\end{equation}
which implies that the virtual wavepacket is created at time $0$, i.e. at the moment when the real wavepacket enters the region with sufficiently strong non-adiabatic coupling. Equation~\eqref{eq:fin_DeltaR2} is valid when such region and the typical energy gap in this region is relatively small and/or that the momentum of the incident (real) wavepacket is high enough, so that the momenta of the real and virtual wavepackets are not very different.
In Section~\ref{sec:algr} we describe a surface hopping algorithm to model non-adiabatic dynamics in molecular systems based on this assumption, e.g. Eqs.~\eqref{eq:fin_DeltaR2}.

\subsection{Detailed balance}

While in the derivation of Eqs.~\eqref{eq:p1_eq2}-\eqref{eq:phifin} we explicitly considered a single nuclear coordinate, it is easy to see that the extension to the multidimensional case is trivial (as long as we neglect the Hessian in Eq.~\eqref{eq:eq_alpha}): $R_{\mathrm{r}(\mathrm{v})}$ and  $\Delta p_{12}$ become multicomponent vectors and therefore the exponents in Eqs.~\eqref{eq:p1_eq2}-\eqref{eq:phifin} contain sums over these components. For the sake of simplicity we will assume that all the degrees of freedom have same mass $M$ and same momentum uncertainty $\sigma^{-1}$.

If the number of the degrees of freedom coupled to the transition is sufficiently high, one can expect that the exponent in the Gaussians in Eqs.~\eqref{eq:p1_eq2}-\eqref{eq:p2_eq2} is sufficiently large when $t-\tau$ is small (i.e., smaller than the typical timescale for the nuclei). In that case Eq.~\eqref{eq:fin_DeltaR1} is applicable, and,
\begin{equation} \label{eq:delta}
 \begin{split}
   \cD(t;\tau) \simeq & \exp{\Big[- \frac{\Delta \mathbf{p}^2_{12}(t-\tau)^2}{4 M^2\sigma^2}\Big]} \\
    &\equiv\exp\left[-\frac{(t-\tau)^2}{\tau_\mathrm{dec}^2} \right]    \,,
 \end{split}
\end{equation}
where
\begin{equation} \label{eq:tau}
\tau_\mathrm{dec}(t) \equiv \frac{2M \sigma(t)}{|\Delta \mathbf{p}_{12}(t)|}\,.
\end{equation}
Parameter $\tau_\mathrm{dec}$ is a timescale at which states of the nuclei (at times $t$ and $t+\tau_\mathrm{dec}$) become (nearly) orthogonal to each other.

When the decoherence time $\tau_\mathrm{dec}$ is sufficiently small, we can replace the argument $\tau$ in the probabilities $\Pop_1(\tau)$ and $\Pop_2(\tau)$ in Eqs.~\eqref{eq:p1_eq2} and \eqref{eq:p2_eq2} by $t$, so that these equations become simple differential (master) equations
\begin{subequations}\label{eq:Markovian}
 \begin{equation}
  \dot{\Pop}_1(t) \approx \Gamma_{21}(t) \Pop_2(t) - \Gamma_{12}(t) \Pop_1(t),
 \end{equation}
 \begin{equation}
  \dot{\Pop}_2(t) \approx \Gamma_{12}(t) \Pop_1(t) - \Gamma_{21}(t) \Pop_2(t).
 \end{equation}
\end{subequations}
The rates $\Gamma_{12}(t)$ and $\Gamma_{21}(t)$ are given by
\begin{equation} \label{eq:Gamma1}
 \begin{split}
   \Gamma_{nn'}(t) = & \, 2 d^2_{12}(t) \int^t_{-\infty} \dd\tau \cos\left[\cE_{nn'}(t)(t-\tau)\right] \\
   & \, \times \exp\left[-\frac{(t-\tau)^2}{\tau_\mathrm{dec}^2} \right],
 \end{split}
\end{equation}
where we have replaced the lower integration limit by $-\infty$ (which is well justified in the strong decoherence limit, i.e.,  when $t\gg\tau_\mathrm{dec}$) and
\begin{equation} \label{eq:Gamma2}
\cE_{nn'}(t)=\Delta E_{nn'}(t) - \frac{\Delta \mathbf{p}^2_{12}(t)}{2M}\,.
\end{equation}
Evaluating Gaussian integral in Eq.~\eqref{eq:Gamma1} we arrive at
\begin{equation}\label{eq:Gamma3}
  \Gamma_{nn'}(t)\approx  \sqrt{\pi} \tau_\mathrm{dec} d^2_{12}(t) \exp\left[-\frac{1}{4}\cE^2_{nn'}(t) \tau^2_\mathrm{dec}\right].
\end{equation}

It is instructive to look at the ratio of the rates,
 \begin{equation} \label{eq:rates_ratio}
 \begin{split}
  \frac{\Gamma_{12}(t)}{\Gamma_{21}(t)} = & \exp\left[- \frac{\Delta \mathbf{p}_{12}^2(t)}{2M} \Delta E_{12}(t) \tau^2_\mathrm{dec}(t)\right]\\
  & = \exp\left[- 2 M \sigma^2 \Delta E_{12}(t)\right]\,.
 \end{split}
 \end{equation}

Equation~\eqref{eq:rates_ratio} is, in fact, the detailed balance condition. To see this recall that $\sigma\sqrt{2}$ is the inverse of the root mean square deviation of momentum, e.g.,  Eqs.~\eqref{eq:GaussW}~-~\eqref{eq:eq_alpha}. Furthermore, if nuclear degrees of freedom are in local equilibrium, the average momentum is zero and therefore
\begin{equation}\nonumber
\frac{1}{2\sigma^2} = \langle p_\alpha^2 \rangle = M T\,,
\end{equation}
where the later equality is due to equipartition condition and $T$ is temperature of the nuclei. Then we have
\begin{equation} \label{eq:rates_ratio2}
\frac{\Gamma_{12}^{\mathrm{eq}}}{\Gamma_{21}^{\mathrm{eq}}} = \exp\left[-\frac{\Delta E_{12}}{T}\right]\,,
 \end{equation}
which is the conventional detailed balance condition. Note that the detailed equilibrium property is a direct consequence of the phase shift $\Delta p_{12}\Delta R/2$ in Eq.~\eqref{eq:phifin}.

\subsection{Nuclear dynamics}
The average value of the nuclear position operator evolves as
\begin{equation} \label{eq:avg_R}
   \partial_t \langle \hat{R}(t) \rangle = - \ii \langle \left[\hat{R}(t), \opH_\mathrm{mj}\right] \rangle = \frac{\langle \hat{p}(t) \rangle}{M}.
\end{equation}

The equation of motion for the nuclear momentum operator reads
\begin{equation} \label{eq:eq_opp}
 \begin{split}
  \partial_t \hat{p}(t) = & \, \sum_{n=1}^2 F_n(R(t)) \hat{\sigma}_{nn} \\
  & \, + \ii \Delta p_{12}(\bar{R}(t))\left[\hat{V}_{12}(t) - \hat{V}_{21}(t)\right].
 \end{split}
\end{equation}
Averaging this equation over the nuclear state Eq.~\eqref{eq:GaussA} leads to
\begin{equation} \label{eq:eq_avgp}
 \partial_t \langle \hat{p}(t) \rangle = \sum_{n=1}^2 F_n(R(t)) \Pop_n(t) + \ii \Delta p_{12}(\bar{R}(t))\dot{\Pop_1}(t).
\end{equation}

Equation~\eqref{eq:eq_avgp} can be viewed as the ``averaged'' equation of motion of the so-called surface hopping dynamics ~\cite{Tully1990}, where nuclei propagating along one energy surface can instantaneously  and randomly hop to another surface with probabilities prescribed by  Eqs.~\eqref{eq:p1_eq2} and ~\eqref{eq:p1_eq2}. To see this, we introduce a discreet random variable $\Sigma$ that can take on values $0$ and $1$ and switches between these values at random times $t_1,\, t_2,\, ...$. We also assume that the switching rate of $\Sigma(t)$ is defined by Eqs.~\eqref{eq:p1_eq2} and ~\eqref{eq:p1_eq2}, so that probabilities  $\Pop_1(t)$ and $\Pop_2(t)=1-\Pop_1(t)$ correspond to the probabilities of $\Sigma$ being equal to $1$ and $0$, respectively, $\langle \Sigma (t)  \rangle = \Pop_1(t)$. Then we define a stochastic equation of motion for the nuclear momentum as,
\begin{equation} \label{eq:eq_opp1}
 \begin{split}
  \partial_t \hat{p}(t) = & \, F_1(R(t)) \Sigma(t) + F_2(R(t))\big[1-\Sigma(t)\big] \\
  & \, + \ii \Delta p_{12}(t)\dot{\Sigma}(t).
 \end{split}
\end{equation}
The first two terms in the rhs of Eq.~\eqref{eq:eq_opp1} describe forces acting on the nuclei depending on the state of the electrons. The second term describes surface hopping. Indeed, since $\Sigma(t)$ in Eq.~\eqref{eq:eq_opp1} changes steplike, momentum changes discontinuously by $\pm\Delta p_{12}(t)$ at random times $t_1,\, t_2,\, ...$, which corresponds to the hopping between the PES.  (Note that the value of the $\Delta p_{12}$ is such that the energy of the nuclei is approximately conserved at the hops.) Upon averaging,   Eq.~\eqref{eq:eq_opp1} transforms into Eq.~\eqref{eq:eq_avgp}, thus the stochastic hops governed by Eq.~\eqref{eq:eq_opp1} on average correspond to the mean dynamics described by  Eq.~\eqref{eq:eq_avgp}.

Note that this statement is correct only in adiabatic basis: In arbitrary basis (e.g. diabatic basis, etc) the last term in the nuclear equation of motion ~\eqref{eq:eq_avgp}  is different, and therefore the hopping description of Eq.~\eqref{eq:eq_opp1} is not applicable.
\section{The algorithm} \label{sec:algr}

In the previous section we derived equations of motion for the propagation of electronic occupation probabilities and nuclear coordinates. The equations for the electronic occupation probabilities significantly simplify when Markovian approximation is applicable as they become simple rate equations. e.g. Eqs.~\eqref{eq:Markovian}. The range of applicability of such Markovian dynamics is limited to the case of strong decoherence, i.e., when many nuclear degrees of freedom are coupled to a given electronic transition so that the decoherence time $\tau_\mathrm{dec}$ is small. This is not necessarily the case, in particular for few-atomic molecules. In such molecules the effects of decoherence are moderate and so one needs to solve more general equations ~\eqref{eq:p1_eq2} and ~\eqref{eq:p1_eq2}.
When the region of non-adiabatic coupling is narrow, one can apply Eq.~\eqref{eq:fin_DeltaR2}, which greatly simplifies the computation.

Furthermore, while realistic non-adiabatic molecular dynamics frequently involves more than two PES, we emphasize that the applicability of the numerical approach to be outlined below is not limited to a two-PES case only. Quite frequently in molecules the non-adiabatic dynamics is limited to the relatively narrow and well separated regions of electronic level crossing. Equations~\eqref{eq:p1_eq2} and ~\eqref{eq:p1_eq2} are specifically designed to treat such regions. After passing through such a region, the nuclear degrees of freedom can again be treated within the conventional Born-Openheimer approximation, until they encounter another level crossing, possibly a different one (or the same), where the non-adiabatic dynamics occurs again. For practical description of such intermittent dynamics the electronic probabilities are being reset after leaving a non-adiabatic region and depending on which electronic state the system is; see below.

In order to model nuclear dynamics based on ~\eqref{eq:p1_eq2},  ~\eqref{eq:p1_eq2}, ~\eqref{eq:eq_opp} and ~\eqref{eq:eq_opp1}, we use the modified FSSH method \cite{Tully1990}. The algorithmic steps are the following:

\begin{enumerate}
 \item We put the nuclear wave packet (referred to as ``real") on the PES $n_0$ at coordinate $\textstyle R_0$ outside the NAC region and launch it towards the latter.
\item When the wave packet reaches the NAC region, point $\tilde{r}$, we spawn the ``virtual" wave packet on the other PES $n'_0$ with momentum $p_\mathrm{r}(t) + \Delta p_{n_0 n'_0}(\tilde{r})$.
 As a criterion for spawning, we use the condition
 \begin{equation} \label{eq:massey}
   \mathcal{M}>\zeta, \quad \mathcal{M} = \left|\frac{\bar{v} A_{12}(\tilde{r})}{\Delta E_{12}(\tilde{r})}\right|,
 \end{equation}
 where $\textstyle \mathcal{M}$ is the Massey parameter\cite{Tully1990}. Typically the value of $\textstyle \zeta$ is much smaller than $1$, in the range of $10^{-2}-10^{-4}$ depending on the problem; see discussion in the next section.
 \item We propagate ``real" and ``virtual" wave packets as classical particles along the trajectories determined by the Newtonian equations of motion:
 \begin{subequations}
  \begin{equation} \label{eq:eq_R}
   \dot{R}_\mathrm{w}(t) = \frac{p_\mathrm{w}(t)}{M},
  \end{equation}
  \begin{equation} \label{eq:eq_p}
   \dot{p}_\mathrm{w}(t) = F_n(R_\mathrm{w}(t)), \quad
  \end{equation}
 \end{subequations}
 where $\mathrm{w} \in \{\mathrm{r}, \mathrm{v}\}$ are indices attributed to ``real" and ``virtual" wave packet, respectively, and $\textstyle n \in \{1, 2\}$ denotes the PES where the wave packet with index $\mathrm{w}$ propagates.
 \item At each time step, we attempt to make a hop of the ``real" wave packet to another PES by checking the condition
 \begin{equation} \label{eq:hop_cond1}
   \frac{\dot{\Pop}_n(t)}{\Pop_n(t)} > \xi,
 \end{equation}
 where $\xi \in \left[0,1\right]$ is a random number.
 \begin{enumerate}
   \item In case, the condition~\eqref{eq:hop_cond1} does not hold, we return to step 3 and continue the propagation of the wave packets along the same PESs using Eqs.~\eqref{eq:eq_R} and \eqref{eq:eq_p}.
   \item If the condition~\eqref{eq:hop_cond1} is fulfilled and the hops of both ``real" and ``virtual" wave packets are energetically allowed, we perform instantaneous hops of both wave packets from their current PES on the opposite PES ($1 \leftrightarrows 2$) and adjust their momenta as
   \begin{equation} \label{eq:adj_p}
     p_\mathrm{w}(t) \overset{n \rightarrow n'}{=}  p_\mathrm{w}(t) - \Delta p_{n n'}(t).
   \end{equation}
   After hops, positions of wave packets remain unaltered.
   Then, we return to step 3 and continue propagation of the wave packets along new PESs using Eqs.~\eqref{eq:eq_R} and \eqref{eq:eq_p} until the next hop event when the condition \eqref{eq:hop_cond1} is met.
 \end{enumerate}
\item When both wave packets leave the region of NAC,.i.e, when $\mathcal{M}<\zeta$, we eliminate the virtual wavepacket and run the real wavepacket according to Eqs.~\eqref{eq:eq_R} and ~\eqref{eq:eq_p} until it reaches another region of NAC or leaves the desired simulation time or space domain. In the former case we proceed with Step 2  and reassign the initial electronic probabilities accordingly (i.e., $\Pop_n=1$ for the real wavepacket). In the latter case the trajectory is complete.  Similarly to the original Tully's approach, to obtain scattering probabilities we run multiple trajectories. Then, we count the number of trajectories of the ``real" wave packet corresponding to a particular scattering scenario (e.g. final PES number and/or forward or backward scattering) and divide it by the total number of simulated trajectories $\textstyle N_\mathrm{trj}$.
\end{enumerate}

The key difference of the proposed approach from the Tully's method is the way we evaluate the populations of the adiabatic states.
For this purpose, we use Eqs.~\eqref{eq:p1_eq2} and \eqref{eq:p2_eq2} which account for the effects of decoherence and phases.
To solve these equations for PESs populations evolution, we propagate a pair of wave packets (``real" and ``virtual") on different PESs, in contrast to a single wave packet used in the original FSSH algorithm.

Now let us discuss in details how the equations of motion for the adiabatic states populations are solved.
Due to the ``decoherence factor" $\cD(t;\tau)$ in  Eqs.~\eqref{eq:p1_eq2} and \eqref{eq:p2_eq2},  the computational complexity of the problem scales $\textstyle \propto~N^2_\mathrm{st}$, where $\textstyle N_\mathrm{st}$ is the number of time steps required to finish the trajectory.
We overcome this issue using the decomposition of the ``decoherence factor", which allows to represent $\textstyle \cD(t;\tau)$ as a series of products of functions dependent on $t$ and $\tau$ (see details in Appendix~\ref{sec:decomp}).
This approach leads to the equations of motion for $\textstyle \Pop_{1,2}(t)$ of the form
\begin{subequations}
 \begin{equation} \label{eq:p1_eq3}
  \begin{split}
   \dot{\Pop}_1(t) \approx & \, 2 d_{12}(t)
   \sum_{j=0}^{\nmax} Y_j(t) \\
   & \, \times \mathrm{Re}\left\{\ee^{\ii \cE_{21}(t)} Z^{(j)}_{21}(t) - \ee^{\ii \cE_{12}(t)} Z^{(j)}_{12}(t)\right\},
  \end{split}
 \end{equation}
 \begin{equation} \label{eq:p2_eq3}
  \begin{split}
   \dot{\Pop}_2(t) \approx & \, 2 d_{12}(t)
   \sum_{j=0}^{\nmax} Y_j(t) \\
   & \, \times \mathrm{Re}\left\{\ee^{\ii \cE_{12}(t)} Z^{(j)}_{12}(t) - \ee^{\ii \cE_{21}(t)} Z^{(j)}_{21}(t)\right\}.
  \end{split}
 \end{equation}
\end{subequations}
Parameter $\textstyle \nmax$ denotes the number of terms we take in the series to ensure the convergence.
Function $\textstyle Y^{(j)}(t)$ is given by
\begin{equation}
 Y^{(j)}(t) = \frac{\lambda^j \sqrt{1-\lambda^2}}{2^j j!} \exp\left[(\lambda-1) \eta^2(t)\right] \mathrm{H}_j(\eta(t)) ,
\end{equation}
where $\eta(t)$ reads as
 \begin{equation}
\eta(t,\tau) = \frac{R_{\mathrm{r}}(t)-R_{\mathrm{v}}(t,\tau)}{2\sigma}\,,
 \end{equation}
 $\textstyle \lambda = (\sqrt{5}-1)/2$, and $\textstyle \mathrm{H}_j(x)$ denotes the $j$-th order Hermite polynomial.
Function $\textstyle Z^{(j)}_{nn'}(t)$ (where $n,n'\in\{1,2\}$ and $n \neq n'$) obeys the equation of motion as follows
\begin{equation}
 \dot{Z}^{(j)}_{nn'}(t) = d_{12}(t) Y^{(j)}(t) \Pop_{n}(t) \ee^{\ii \cE_{nn'}(t)}
\end{equation}
with the initial condition $\textstyle Z^{(j)}_{nn'}(0) = 0$.

\section{Numerical results and Discussion} \label{sec:num}

We test our approach on a set of test problems proposed by J.~Tully.~\cite{Tully1990}.
This set includes three model problems involving one nuclear degree of freedom and a pair of coupled PESs and is routinely employed for verification of novel NAMD methods.
The details of these problems are presented in Appendix~\ref{sec:tully}.
We compare the results obtained using the approach presented in this paper to the results obtained via the exact numerical solution of the TDSE and the standard FSSH method.
In all three problems, the wave packet initially resides on the lower PES and its wavefunction is given by
\begin{equation}
 |\Gs_1(R_0, p_0; R,0)\rangle = \frac{1}{\sqrt[4]{\pi \sigma_0^2}} \exp\left[\ii p_0 R - \frac{(R-R_0)^2}{2\sigma^2_0}\right]|1\rangle,
\end{equation}
where the wave packet initial position $R_0 < 0$ is set outside the NAC region, and $p_0 > 0$ stands for the nuclear initial momentum.
The initial width of the wave packet is taken to be $\sigma_0 = 20/p_0$ as in Tully~\cite{Tully1990}.
For all problems, the nuclear mass is set close to the mass of a hydrogen nucleus (proton),  $\textstyle M=2000~\mathrm{a.u}$.
For computation, we set $N_\mathrm{nt} = 5$.
To calculate the scattering probabilities, we sampled $N_\mathrm{trj}=2500$ trajectories for each value of the initial momentum $p_0$ for both our approach and the FSSH method.
In what follows all quantities are given in the atomic units.

\begin{figure}[t!] 
	\centering
	\includegraphics{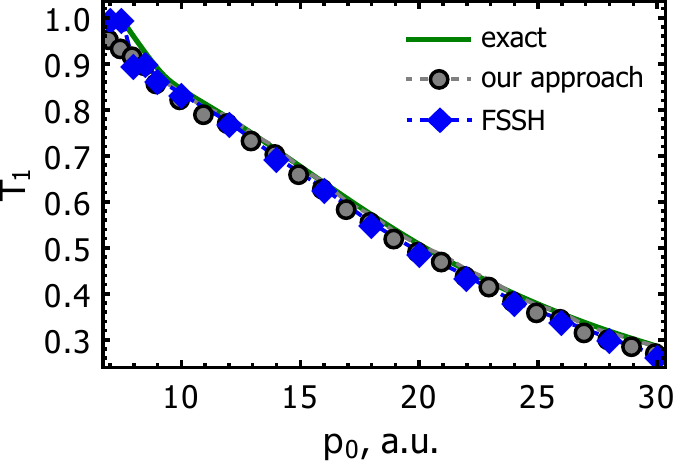}
	\caption{The transmission probability of the nuclear wave packet on the lower PES versus the initial momentum $p_0$ for the SAC problem. \label{fig:fig_sac}}
\end{figure}

\begin{figure}[b!] 
	\centering
	\includegraphics{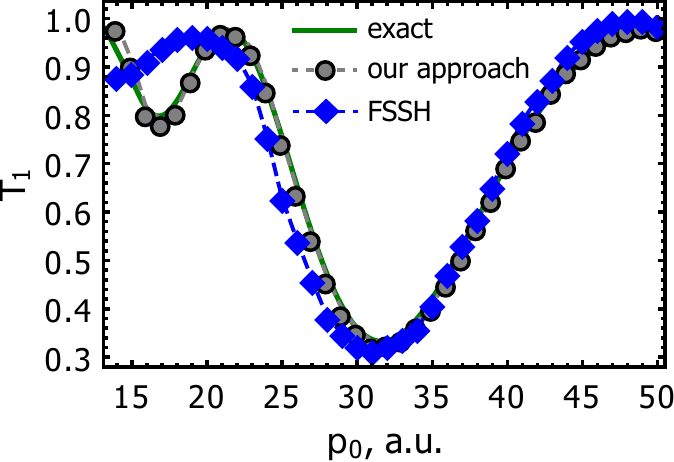}
	\caption{The transmission probability of the nuclear wave packet on the lower PES versus the initial momentum $p_0$ for the DAC problem. \label{fig:fig_dac}}
\end{figure}

The first problem we consider is the single avoided crossing (SAC).
For this problem, we choose $\zeta = 10^{-4}$ in the condition~\eqref{eq:massey}.
Figure~\ref{fig:fig_sac} demonstrates the dependence of the probability of the wave packet transmission on the lower PES $\textstyle T_1$ on the initial momentum calculated using the TDSE, FSSH, and our approach.
Computations demonstrate that for this problem, both the standard FSSH and our approach exhibit good agreement with the exact results.

The second model, a double avoided crossing (DAC), features a quantum interference between two pathways along lower and upper PESs, which results in the Stueckelberg oscillations \cite{Stueckelberg1932} in the scattering probability as shown in Fig.~\ref{fig:fig_dac}.
The FSSH method works well for large values of the initial momentum $p_0>30$.
However, for the lower $p_0$ the results given by the FSSH and the exact results are out of phase.
Our approach reproduces the exact results quantitatively for $p_0>20$ and qualitatively for $p_0<20$.
This implies that our approach correctly grasps the interference effects since it operates with two wave packets rather than with one as in the standard FSSH approach. Note that Eqs.~\eqref{eq:p1_eq2} and \eqref{eq:p2_eq2} account for different positions of real and virtual wavepackets, which leads, in particular, to the extra phase shift term in Eq.\eqref{eq:phifin}. We find $\zeta = 10^{-2}$ in Eq.~\eqref{eq:massey} to be optimal for this problem.

\begin{figure*}[t!] 
	\centering
	\includegraphics{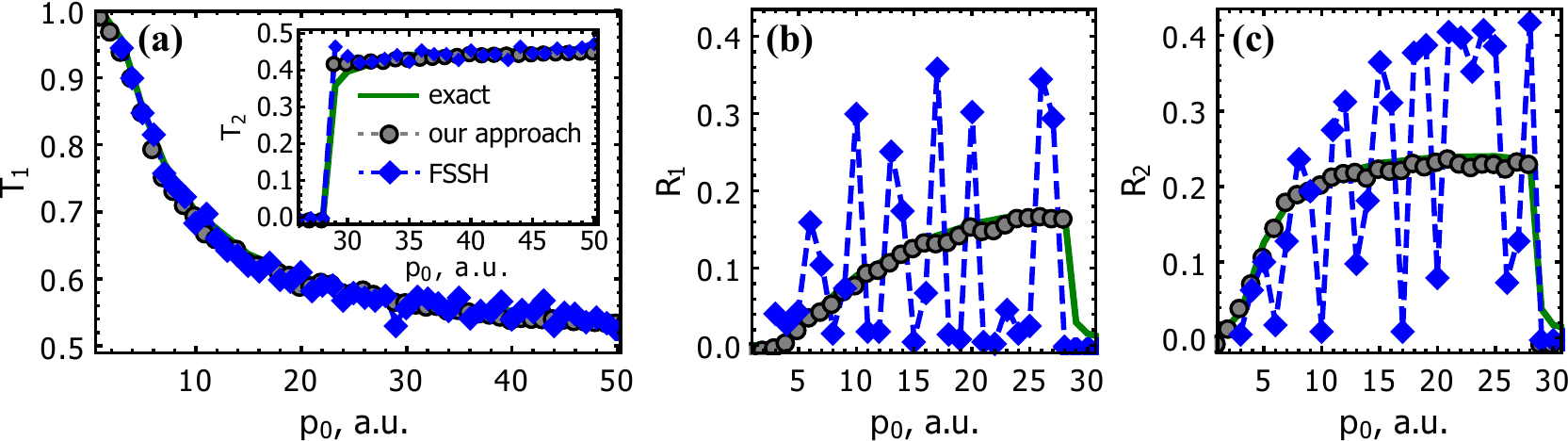}
	\caption{Scattering probabilities versus the initial wave packet momentum for the ECR problem: (a) probability of transmission on the lower PES, inset demonstrates the transmission probability on the upper PES; (b) reflection probability on the lower PES; (c) reflection probability on the upper PES. \label{fig:fig_ecr}}
\end{figure*}

The third problem we use to test our approach is frequently referred to as an extended coupling with reflection (ECR).
It is distinctive by the fact that for the initial wave packet momenta $p_0 < 28$ the conventional FSSH method is incapable of reproducing the results given by the TDSE quantitatively nor qualitatively (see Figs.~\ref{fig:fig_ecr}b and \ref{fig:fig_ecr}c).
This failure occurs due to the lack of decoherence in the standard FSSH algorithm \cite{Schwartz1996, Jaeger2012, Subotnik2010, Subotnik2011, Shenvi2011, Shenvi2012, Subotnik2013, Akimov2014}.
Note that our equations for the adiabatic states populations, e.g., Eqs.~\eqref{eq:p1_eq2} and \eqref{eq:p2_eq2}, do not account for the scenario, when one of the wave packets used for simulation propagates along the upper PES and reflects off the potential energy barrier, thus changing its direction of propagation. This situation, however, is taken care of by the spawning procedure: When wave packets leave the NAC region, which is determined from the condition $\mathcal{M}<\zeta$ (with $\zeta=10^{-3}$ for this problem), the virtual wavepacket is eliminated and probabilities are reset as described in Step 5 of the algorithm. Then the real wavepacket either leaves the computation domain (if its on the lower PES), or returns back to the NAC region (if it is on the upper PES), where the computation proceeds from Step 2. As demonstrated by Fig.~\ref{fig:fig_ecr}, our method reproduces the exact numerical results.

The computational cost of our algorithm is somewhat higher compared to the standard FSSH algorithm since it is required to solve more differential equations. However, this is a reasonable trade-off for much more consistent results in DAC and ECR problems.

In summary, in this paper we developed a formalism to describe the decoherence effects related to quantum fluctuations of nuclear positions (and momenta). We have shown that the proper account of superpositions of the nuclear wavefunctions corresponding to the classical nuclear trajectories along different PES leads to the detailed balance property for the electronic populations. Also, using this formalism, we have modified the FSSH algorithm to account for the aforementioned decoherence effects.

\begin{acknowledgements}
We thank Roman Baskov and Sergey Tretiak for valuable discussions. This work was performed under the auspices of the U.S. Department of Energy under Contract No. 89233218CNA000001 and was supported in part by the U.S. Department of Energy LDRD program at Los Alamos National Laboratory through the grant No.20200074ER.
\end{acknowledgements}

\section*{Data Availability}

The data that supports the findings of this study are available within the article.

\appendix

\begin{figure*}[t!] 
	\centering
	\includegraphics[width=\textwidth]{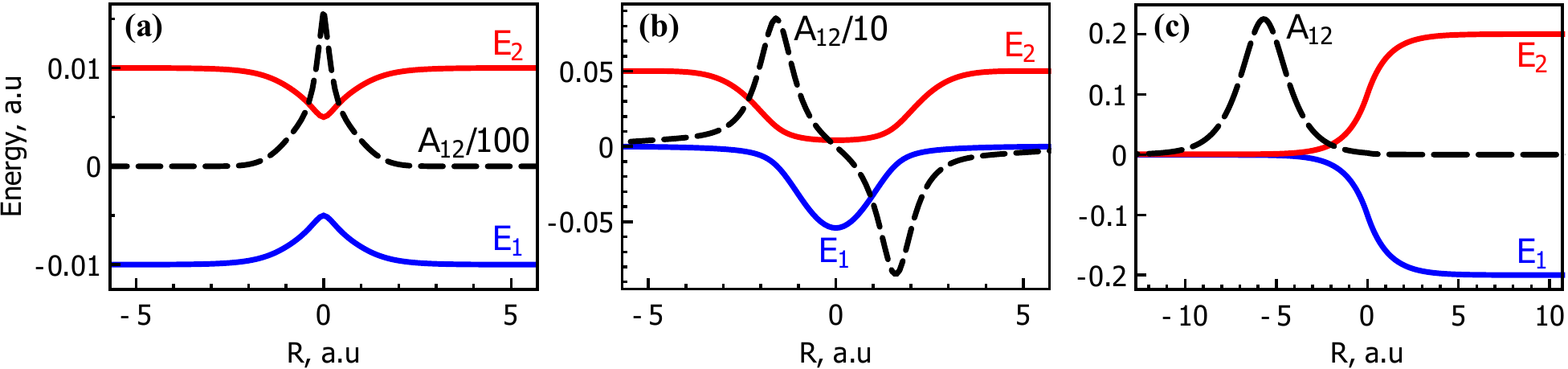}
	\caption{Tully's problem set: (a) single avoided crossing, (b) double avoided crossing, and (c) extended coupling with reflection. Solid lines represent the adiabatic PESs. Black dashed lines show the non-adiabatic couplings.
		\label{fig:fig_tully}}
\end{figure*}

\section{The Hamiltonian in the ``velocity gauge"} \label{sec:hvg_der}
Using that the adiabatic electronic states form a complete orthonormal basis
\begin{subequations}
 \begin{equation}  \label{eq:unity_exp}
  \sum_{n} |n(\vecR)\rangle\langle n(\vecR)| = 1,
 \end{equation}
 \begin{equation}  \label{eq:orth}
  \langle n(\vecR)|n'(\vecR)\rangle = \delta_{nn'},
 \end{equation}
\end{subequations}
one can formally represent the Hamiltonian \eqref{eq:full_ham} as
\begin{equation}
 \opH = \sum_{n}\sum_{n'} \langle n(\vecR)|\opH|n'(\vecR)\rangle |n(\vecR)\rangle\langle n'(\vecR)|,
\end{equation}
where $\textstyle \langle n(\vecR)|\opH|n'(\vecR)\rangle$ reads
\begin{equation} \label{eq:ham_expand}
 \langle n(\vecR)|\opH|n'(\vecR)\rangle = -\sum_{\alpha} \frac{\langle n(\vecR)|\nabla^2_\vecR|n'(\vecR)\rangle}{2M_\alpha} + E_n(\vecR) \delta_{nn'}.
\end{equation}
The matrix elements $\textstyle \left\langle n(\vecR)\left|\nabla^2_\vecR\right|n'(\vecR)\right\rangle$ can be expressed as \cite{Baer2002, Baer2006, Gherib2016, Baskov2019}
\begin{equation}
 \begin{split}
  \left\langle n(\vecR)\left|\nabla^2_\vecR\right|n'(\vecR)\right\rangle = & \, \delta_{nn'} \nabla^2_\vecR \\
  & \, + \langle n(\vecR)|\nabla_\vecR n'(\vecR)\rangle \nabla_\vecR \\
  & \, + \nabla_\vecR \langle n(\vecR)|\nabla_\vecR n'(\vecR)\rangle \\
  & \, - \langle \nabla_\vecR n(\vecR)|\nabla_\vecR n'(\vecR)\rangle.
 \end{split}
\end{equation}
Inserting the resolution of unity \eqref{eq:unity_exp} into the last term on the rhs of the above expression and recalling that $\textstyle \langle n(\vecR)|\nabla_\vecR n(\vecR)\rangle = \vecA_{nn'}(\vecR)$, one obtains
\begin{equation}
\begin{split}
 \left\langle n(\vecR)\left|\nabla^2_\vecR\right|n'(\vecR)\right\rangle =
  & \, \delta_{nn'} \nabla^2_\vecR - \vecA^2_{nn'} \\
  & \, + \vecA_{nn'} \nabla_\vecR + \nabla_\vecR \vecA_{nn'} .
 \end{split}
\end{equation}
Using this result in Eq.~\eqref{eq:ham_expand}, one arrives at the effective molecular Hamiltonian in a ``velocity gauge":\cite{Baer2002,Baer2006,Jasper2004,Takatsuka2007}
\begin{equation} \label{eq:ham_lg}
 \opH_\mathrm{vg} = \sum_{\alpha} \frac{\left(\opp_\alpha - \ii \opA_\alpha\right)^2}{2M_\alpha} + \sum_{n} E_n(\vecR)|n(\vecR)\rangle\langle n(\vecR)|,
\end{equation}
where $\textstyle \opA = \sum_{n,n'}\vecA_{nn'}(\vecR)|n(\vecR)\rangle\langle n'(\vecR)|$.

\section{Decomposition of the ``decoherence factor" $\textstyle \cD(t;\tau)$} \label{sec:decomp}
Let us formally rewrite $\textstyle \cD(t;\tau)$ given by Eq.~\eqref{eq:D} as
\begin{equation} \label{eq:rewr_D}
 \cD(t;\tau) = \exp\left\{\kappa \alpha^2 [\eta^2(t) + \eta^2(\tau)]\right\} W(t;\tau)
\end{equation}
where $\kappa = \lambda - 1/2$ and $\textstyle W(t;\tau)$ has the form
\begin{equation} \label{eq:decomp}
 \begin{split}
     W(t;\tau) = & \, \exp\left\{-\alpha^2 \frac{1+\lambda^2}{2(1-\lambda^2)}\left[\eta^2(t) + \eta^2(\tau)\right]\right\} \\
     & \, \times \exp\left[2\alpha^2 \frac{\lambda}{1-\lambda^2} \eta(t) \eta(\tau)\right].
 \end{split}
\end{equation}
Function $\textstyle W(t;\tau)$ can be decomposed as\cite{URen2003}
\begin{equation}
  W(t; \tau) = \sqrt{1-\lambda^2} \sum_{j=0}^\infty \lambda^j \varTheta_j(\alpha \eta(t)) \varTheta_j(\alpha \eta(\tau)),
\end{equation}
where $\textstyle \varTheta_j(x)$ is given by
\begin{equation}
 \varTheta_j(x) = \sqrt{\frac{1}{2^j j!}} \, \mathrm{H}_j(x) \ee^{-x^2/2}.
\end{equation}
Combining Eq.~\eqref{eq:decomp} with Eq.~\eqref{eq:rewr_D}, substituting the result into Eq.~\eqref{eq:p1_eq2} and limiting the number of terms in the decomposition to $\textstyle N_\mathrm{nt}$, one obtains Eq.~\eqref{eq:p1_eq3}.
\section{Tully's problem suite} \label{sec:tully}

For the two-state problems with single nuclear degree of freedom, the molecular Hamiltonian in the diabatic representation acquires the form
\begin{equation}
 \opH = - \frac{\partial_R^2}{2M} \, \hat{\mathrm{I}} +
 \begin{pmatrix}
  H^\mathrm{el}_{11}(R) && H^\mathrm{el}_{12}(R) \\
  H^\mathrm{el}_{21}(R) && H^\mathrm{el}_{22}(R)
 \end{pmatrix},
\end{equation}
where $\textstyle \hat{\mathrm{I}}$ is a $\textstyle 2\times2$ identity matrix and the matrix $H^\mathrm{el}$ determines a pair of coupled PES in the diabatic representation.

For the SAC problem, the elements of the matrix $H^\mathrm{el}$ are given by
\begin{equation}
 \begin{split}
  H^\mathrm{el}_{11}(R) &= 0.01\mathrm{sgn}(R)\left(1-\ee^{-1.6|R|}\right), \\
  H^\mathrm{el}_{22}(R) &= - H^\mathrm{el}_{11}(R), \\
  H^\mathrm{el}_{12}(R) &= H^\mathrm{el}_{21}(R) = 0.005 \ee^{-R^2}.
 \end{split}
\end{equation}
The corresponding adiabatic PESs and NACs are shown in Fig.~\ref{fig:fig_tully}a.
For the DAC problem, the matrix $H^\mathrm{el}$ is determined as
\begin{equation}
 \begin{split}
  H^\mathrm{el}_{11}(R) & = 0, \\
  H^\mathrm{el}_{22}(R) & = 0.05 - 0.1 \ee^{-0.28R^2}, \\
  H^\mathrm{el}_{12}(R) & = H^\mathrm{el}_{21}(R) = 0.015 \ee^{-0.06 R^2}.
 \end{split}
\end{equation}
Figure~\ref{fig:fig_tully}b shows the corresponding adiabatic PESs and NACs.
The ECR problem is given by the following diabatic surfaces
\begin{equation} \label{eq:ecr_dia}
 \begin{split}
  H^\mathrm{el}_{11}(R) = & \, 6 \times 10^{-4}, \\
  H^\mathrm{el}_{22}(R) = & \, - H^\mathrm{el}_{11}(R)
 \end{split}
\end{equation}
and couplings
 \begin{equation} \label{eq:ecr_coupl}
  H^\mathrm{el}_{12}(R) =
  \begin{cases}
    0.1 \ee^{0.9R}, \quad R<0, \\
    0.1 (2-\ee^{-0.9R}), \quad R\ge 0.
  \end{cases}
\end{equation}
The corresponding adiabatic PESs and NACs are demonstrated in Fig.~\ref{fig:fig_tully}c.
\bibliography{bibliography}

\begin{thebibliography}{46}%
\makeatletter
\providecommand \@ifxundefined [1]{%
 \@ifx{#1\undefined}
}%
\providecommand \@ifnum [1]{%
 \ifnum #1\expandafter \@firstoftwo
 \else \expandafter \@secondoftwo
 \fi
}%
\providecommand \@ifx [1]{%
 \ifx #1\expandafter \@firstoftwo
 \else \expandafter \@secondoftwo
 \fi
}%
\providecommand \natexlab [1]{#1}%
\providecommand \enquote  [1]{``#1''}%
\providecommand \bibnamefont  [1]{#1}%
\providecommand \bibfnamefont [1]{#1}%
\providecommand \citenamefont [1]{#1}%
\providecommand \href@noop [0]{\@secondoftwo}%
\providecommand \href [0]{\begingroup \@sanitize@url \@href}%
\providecommand \@href[1]{\@@startlink{#1}\@@href}%
\providecommand \@@href[1]{\endgroup#1\@@endlink}%
\providecommand \@sanitize@url [0]{\catcode `\\12\catcode `\$12\catcode
  `\&12\catcode `\#12\catcode `\^12\catcode `\_12\catcode `\%12\relax}%
\providecommand \@@startlink[1]{}%
\providecommand \@@endlink[0]{}%
\providecommand \url  [0]{\begingroup\@sanitize@url \@url }%
\providecommand \@url [1]{\endgroup\@href {#1}{\urlprefix }}%
\providecommand \urlprefix  [0]{URL }%
\providecommand \Eprint [0]{\href }%
\providecommand \doibase [0]{https://doi.org/}%
\providecommand \selectlanguage [0]{\@gobble}%
\providecommand \bibinfo  [0]{\@secondoftwo}%
\providecommand \bibfield  [0]{\@secondoftwo}%
\providecommand \translation [1]{[#1]}%
\providecommand \BibitemOpen [0]{}%
\providecommand \bibitemStop [0]{}%
\providecommand \bibitemNoStop [0]{.\EOS\space}%
\providecommand \EOS [0]{\spacefactor3000\relax}%
\providecommand \BibitemShut  [1]{\csname bibitem#1\endcsname}%
\let\auto@bib@innerbib\@empty
\bibitem [{\citenamefont {Nelson}\ \emph {et~al.}(2020)\citenamefont {Nelson},
  \citenamefont {White}, \citenamefont {Bjorgaard}, \citenamefont {Sifain},
  \citenamefont {Zhang}, \citenamefont {Nebgen}, \citenamefont
  {Fernandez-Alberti}, \citenamefont {Mozyrsky}, \citenamefont {Roitberg},\
  and\ \citenamefont {Tretiak}}]{Nelson20}%
  \BibitemOpen
  \bibfield  {author} {\bibinfo {author} {\bibfnamefont {T.~R.}\ \bibnamefont
  {Nelson}}, \bibinfo {author} {\bibfnamefont {A.~J.}\ \bibnamefont {White}},
  \bibinfo {author} {\bibfnamefont {J.~A.}\ \bibnamefont {Bjorgaard}}, \bibinfo
  {author} {\bibfnamefont {A.~E.}\ \bibnamefont {Sifain}}, \bibinfo {author}
  {\bibfnamefont {Y.}~\bibnamefont {Zhang}}, \bibinfo {author} {\bibfnamefont
  {B.}~\bibnamefont {Nebgen}}, \bibinfo {author} {\bibfnamefont
  {S.}~\bibnamefont {Fernandez-Alberti}}, \bibinfo {author} {\bibfnamefont
  {D.}~\bibnamefont {Mozyrsky}}, \bibinfo {author} {\bibfnamefont {A.~E.}\
  \bibnamefont {Roitberg}},\ and\ \bibinfo {author} {\bibfnamefont
  {S.}~\bibnamefont {Tretiak}},\ }\bibfield  {title} {\enquote {\bibinfo
  {title} {Non-adiabatic excited-state molecular dynamics: Theory and
  applications for modeling photophysics in extended molecular materials},}\
  }\href {https://doi.org/10.1021/acs.chemrev.9b00447} {\bibfield  {journal}
  {\bibinfo  {journal} {Chemical Reviews}\ }\textbf {\bibinfo {volume} {120}},\
  \bibinfo {pages} {2215--2287} (\bibinfo {year} {2020})},\ \bibinfo {note}
  {pMID: 32040312},\ \Eprint
  {https://arxiv.org/abs/https://doi.org/10.1021/acs.chemrev.9b00447}
  {https://doi.org/10.1021/acs.chemrev.9b00447} \BibitemShut {NoStop}%
\bibitem [{\citenamefont {C.~Tully}(1998)}]{Tully98}%
  \BibitemOpen
  \bibfield  {author} {\bibinfo {author} {\bibfnamefont {J.}~\bibnamefont
  {C.~Tully}},\ }\bibfield  {title} {\enquote {\bibinfo {title} {Mixed
  quantum--classical dynamics},}\ }\href {https://doi.org/10.1039/A801824C}
  {\bibfield  {journal} {\bibinfo  {journal} {Faraday Discuss.}\ }\textbf
  {\bibinfo {volume} {110}},\ \bibinfo {pages} {407--419} (\bibinfo {year}
  {1998})}\BibitemShut {NoStop}%
\bibitem [{\citenamefont {Subotnik}\ \emph {et~al.}(2016)\citenamefont
  {Subotnik}, \citenamefont {Jain}, \citenamefont {Landry}, \citenamefont
  {Petit}, \citenamefont {Ouyang},\ and\ \citenamefont
  {Bellonzi}}]{Subotnik16}%
  \BibitemOpen
  \bibfield  {author} {\bibinfo {author} {\bibfnamefont {J.~E.}\ \bibnamefont
  {Subotnik}}, \bibinfo {author} {\bibfnamefont {A.}~\bibnamefont {Jain}},
  \bibinfo {author} {\bibfnamefont {B.}~\bibnamefont {Landry}}, \bibinfo
  {author} {\bibfnamefont {A.}~\bibnamefont {Petit}}, \bibinfo {author}
  {\bibfnamefont {W.}~\bibnamefont {Ouyang}},\ and\ \bibinfo {author}
  {\bibfnamefont {N.}~\bibnamefont {Bellonzi}},\ }\bibfield  {title} {\enquote
  {\bibinfo {title} {Understanding the surface hopping view of electronic
  transitions and decoherence},}\ }\href
  {https://doi.org/10.1146/annurev-physchem-040215-112245} {\bibfield
  {journal} {\bibinfo  {journal} {Annual Review of Physical Chemistry}\
  }\textbf {\bibinfo {volume} {67}},\ \bibinfo {pages} {387--417} (\bibinfo
  {year} {2016})},\ \bibinfo {note} {pMID: 27215818},\ \Eprint
  {https://arxiv.org/abs/https://doi.org/10.1146/annurev-physchem-040215-112245}
  {https://doi.org/10.1146/annurev-physchem-040215-112245} \BibitemShut
  {NoStop}%
\bibitem [{\citenamefont {Crespo-Otero}\ and\ \citenamefont
  {Barbatti}(2018)}]{Barbatti18}%
  \BibitemOpen
  \bibfield  {author} {\bibinfo {author} {\bibfnamefont {R.}~\bibnamefont
  {Crespo-Otero}}\ and\ \bibinfo {author} {\bibfnamefont {M.}~\bibnamefont
  {Barbatti}},\ }\bibfield  {title} {\enquote {\bibinfo {title} {Recent
  advances and perspectives on nonadiabatic mixed quantum--classical
  dynamics},}\ }\href {https://doi.org/10.1021/acs.chemrev.7b00577} {\bibfield
  {journal} {\bibinfo  {journal} {Chemical Reviews}\ }\textbf {\bibinfo
  {volume} {118}},\ \bibinfo {pages} {7026--7068} (\bibinfo {year} {2018})},\
  \bibinfo {note} {pMID: 29767966},\ \Eprint
  {https://arxiv.org/abs/https://doi.org/10.1021/acs.chemrev.7b00577}
  {https://doi.org/10.1021/acs.chemrev.7b00577} \BibitemShut {NoStop}%
\bibitem [{\citenamefont {Shalashilin}\ and\ \citenamefont
  {Child}(2005)}]{Shalashilin05}%
  \BibitemOpen
  \bibfield  {author} {\bibinfo {author} {\bibfnamefont {D.~V.}\ \bibnamefont
  {Shalashilin}}\ and\ \bibinfo {author} {\bibfnamefont {M.~S.}\ \bibnamefont
  {Child}},\ }\bibfield  {title} {\enquote {\bibinfo {title} {Electronic energy
  levels with the help of trajectory-guided random grid of coupled wave
  packets. {I.} {S}ix-dimensional simulation of {H2}},}\ }\href
  {https://doi.org/10.1063/1.1926268} {\bibfield  {journal} {\bibinfo
  {journal} {The Journal of Chemical Physics}\ }\textbf {\bibinfo {volume}
  {122}},\ \bibinfo {pages} {224108} (\bibinfo {year} {2005})},\ \Eprint
  {https://arxiv.org/abs/https://doi.org/10.1063/1.1926268}
  {https://doi.org/10.1063/1.1926268} \BibitemShut {NoStop}%
\bibitem [{\citenamefont {Shalashilin}\ and\ \citenamefont
  {Burghardt}(2008)}]{Shalashilin08}%
  \BibitemOpen
  \bibfield  {author} {\bibinfo {author} {\bibfnamefont {D.~V.}\ \bibnamefont
  {Shalashilin}}\ and\ \bibinfo {author} {\bibfnamefont {I.}~\bibnamefont
  {Burghardt}},\ }\bibfield  {title} {\enquote {\bibinfo {title}
  {Gaussian-based techniques for quantum propagation from the time-dependent
  variational principle: Formulation in terms of trajectories of coupled
  classical and quantum variables},}\ }\href
  {https://doi.org/10.1063/1.2969101} {\bibfield  {journal} {\bibinfo
  {journal} {The Journal of Chemical Physics}\ }\textbf {\bibinfo {volume}
  {129}},\ \bibinfo {pages} {084104} (\bibinfo {year} {2008})},\ \Eprint
  {https://arxiv.org/abs/https://doi.org/10.1063/1.2969101}
  {https://doi.org/10.1063/1.2969101} \BibitemShut {NoStop}%
\bibitem [{\citenamefont {Habershon}(2012)}]{Habershon12}%
  \BibitemOpen
  \bibfield  {author} {\bibinfo {author} {\bibfnamefont {S.}~\bibnamefont
  {Habershon}},\ }\bibfield  {title} {\enquote {\bibinfo {title}
  {Trajectory-guided configuration interaction simulations of multidimensional
  quantum dynamics},}\ }\href {https://doi.org/10.1063/1.3681167} {\bibfield
  {journal} {\bibinfo  {journal} {The Journal of Chemical Physics}\ }\textbf
  {\bibinfo {volume} {136}},\ \bibinfo {pages} {054109} (\bibinfo {year}
  {2012})},\ \Eprint {https://arxiv.org/abs/https://doi.org/10.1063/1.3681167}
  {https://doi.org/10.1063/1.3681167} \BibitemShut {NoStop}%
\bibitem [{\citenamefont {Saller}\ and\ \citenamefont
  {Habershon}(2015)}]{Habershon15}%
  \BibitemOpen
  \bibfield  {author} {\bibinfo {author} {\bibfnamefont {M.~A.~C.}\
  \bibnamefont {Saller}}\ and\ \bibinfo {author} {\bibfnamefont
  {S.}~\bibnamefont {Habershon}},\ }\bibfield  {title} {\enquote {\bibinfo
  {title} {Basis set generation for quantum dynamics simulations using simple
  trajectory-based methods},}\ }\href {https://doi.org/10.1021/ct500657f}
  {\bibfield  {journal} {\bibinfo  {journal} {Journal of Chemical Theory and
  Computation}\ }\textbf {\bibinfo {volume} {11}},\ \bibinfo {pages} {8--16}
  (\bibinfo {year} {2015})},\ \bibinfo {note} {pMID: 26574198},\ \Eprint
  {https://arxiv.org/abs/https://doi.org/10.1021/ct500657f}
  {https://doi.org/10.1021/ct500657f} \BibitemShut {NoStop}%
\bibitem [{\citenamefont {Grigolo}, \citenamefont {Viscondi},\ and\
  \citenamefont {de~Aguiar}(2016)}]{Grigolo16}%
  \BibitemOpen
  \bibfield  {author} {\bibinfo {author} {\bibfnamefont {A.}~\bibnamefont
  {Grigolo}}, \bibinfo {author} {\bibfnamefont {T.~F.}\ \bibnamefont
  {Viscondi}},\ and\ \bibinfo {author} {\bibfnamefont {M.~A.~M.}\ \bibnamefont
  {de~Aguiar}},\ }\bibfield  {title} {\enquote {\bibinfo {title}
  {Multiconfigurational quantum propagation with trajectory-guided generalized
  coherent states},}\ }\href {https://doi.org/10.1063/1.4942926} {\bibfield
  {journal} {\bibinfo  {journal} {The Journal of Chemical Physics}\ }\textbf
  {\bibinfo {volume} {144}},\ \bibinfo {pages} {094106} (\bibinfo {year}
  {2016})},\ \Eprint {https://arxiv.org/abs/https://doi.org/10.1063/1.4942926}
  {https://doi.org/10.1063/1.4942926} \BibitemShut {NoStop}%
\bibitem [{\citenamefont {Makhov}\ \emph {et~al.}(2017)\citenamefont {Makhov},
  \citenamefont {Symonds}, \citenamefont {Fernandez-Alberti},\ and\
  \citenamefont {Shalashilin}}]{MAKHOV2017200}%
  \BibitemOpen
  \bibfield  {author} {\bibinfo {author} {\bibfnamefont {D.~V.}\ \bibnamefont
  {Makhov}}, \bibinfo {author} {\bibfnamefont {C.}~\bibnamefont {Symonds}},
  \bibinfo {author} {\bibfnamefont {S.}~\bibnamefont {Fernandez-Alberti}},\
  and\ \bibinfo {author} {\bibfnamefont {D.~V.}\ \bibnamefont {Shalashilin}},\
  }\bibfield  {title} {\enquote {\bibinfo {title} {Ab initio quantum direct
  dynamics simulations of ultrafast photochemistry with multiconfigurational
  {E}hrenfest approach},}\ }\href
  {https://doi.org/https://doi.org/10.1016/j.chemphys.2017.04.003} {\bibfield
  {journal} {\bibinfo  {journal} {Chemical Physics}\ }\textbf {\bibinfo
  {volume} {493}},\ \bibinfo {pages} {200 -- 218} (\bibinfo {year}
  {2017})}\BibitemShut {NoStop}%
\bibitem [{\citenamefont {Shalashilin}(2009)}]{Shalashilin09}%
  \BibitemOpen
  \bibfield  {author} {\bibinfo {author} {\bibfnamefont {D.~V.}\ \bibnamefont
  {Shalashilin}},\ }\bibfield  {title} {\enquote {\bibinfo {title} {Quantum
  mechanics with the basis set guided by {E}hrenfest trajectories: Theory and
  application to spin-boson model},}\ }\href
  {https://doi.org/10.1063/1.3153302} {\bibfield  {journal} {\bibinfo
  {journal} {The Journal of Chemical Physics}\ }\textbf {\bibinfo {volume}
  {130}},\ \bibinfo {pages} {244101} (\bibinfo {year} {2009})},\ \Eprint
  {https://arxiv.org/abs/https://doi.org/10.1063/1.3153302}
  {https://doi.org/10.1063/1.3153302} \BibitemShut {NoStop}%
\bibitem [{\citenamefont {Shalashilin}(2010)}]{Shalashilin10}%
  \BibitemOpen
  \bibfield  {author} {\bibinfo {author} {\bibfnamefont {D.~V.}\ \bibnamefont
  {Shalashilin}},\ }\bibfield  {title} {\enquote {\bibinfo {title}
  {Nonadiabatic dynamics with the help of multiconfigurational {E}hrenfest
  method: Improved theory and fully quantum 24{D} simulation of pyrazine},}\
  }\href {https://doi.org/10.1063/1.3442747} {\bibfield  {journal} {\bibinfo
  {journal} {The Journal of Chemical Physics}\ }\textbf {\bibinfo {volume}
  {132}},\ \bibinfo {pages} {244111} (\bibinfo {year} {2010})},\ \Eprint
  {https://arxiv.org/abs/https://doi.org/10.1063/1.3442747}
  {https://doi.org/10.1063/1.3442747} \BibitemShut {NoStop}%
\bibitem [{\citenamefont {Makhov}\ \emph {et~al.}(2014)\citenamefont {Makhov},
  \citenamefont {Glover}, \citenamefont {Martinez},\ and\ \citenamefont
  {Shalashilin}}]{Makhov14}%
  \BibitemOpen
  \bibfield  {author} {\bibinfo {author} {\bibfnamefont {D.~V.}\ \bibnamefont
  {Makhov}}, \bibinfo {author} {\bibfnamefont {W.~J.}\ \bibnamefont {Glover}},
  \bibinfo {author} {\bibfnamefont {T.~J.}\ \bibnamefont {Martinez}},\ and\
  \bibinfo {author} {\bibfnamefont {D.~V.}\ \bibnamefont {Shalashilin}},\
  }\bibfield  {title} {\enquote {\bibinfo {title} {Ab initio multiple cloning
  algorithm for quantum nonadiabatic molecular dynamics},}\ }\href
  {https://doi.org/10.1063/1.4891530} {\bibfield  {journal} {\bibinfo
  {journal} {The Journal of Chemical Physics}\ }\textbf {\bibinfo {volume}
  {141}},\ \bibinfo {pages} {054110} (\bibinfo {year} {2014})},\ \Eprint
  {https://arxiv.org/abs/https://doi.org/10.1063/1.4891530}
  {https://doi.org/10.1063/1.4891530} \BibitemShut {NoStop}%
\bibitem [{\citenamefont {Yang}\ \emph {et~al.}(2009)\citenamefont {Yang},
  \citenamefont {Coe}, \citenamefont {Kaduk},\ and\ \citenamefont
  {Mart{\'\i}nez}}]{Yang09}%
  \BibitemOpen
  \bibfield  {author} {\bibinfo {author} {\bibfnamefont {S.}~\bibnamefont
  {Yang}}, \bibinfo {author} {\bibfnamefont {J.~D.}\ \bibnamefont {Coe}},
  \bibinfo {author} {\bibfnamefont {B.}~\bibnamefont {Kaduk}},\ and\ \bibinfo
  {author} {\bibfnamefont {T.~J.}\ \bibnamefont {Mart{\'\i}nez}},\ }\bibfield
  {title} {\enquote {\bibinfo {title} {An ``optimal'' spawning algorithm for
  adaptive basis set expansion in nonadiabatic dynamics},}\ }\href
  {https://doi.org/10.1063/1.3103930} {\bibfield  {journal} {\bibinfo
  {journal} {The Journal of Chemical Physics}\ }\textbf {\bibinfo {volume}
  {130}},\ \bibinfo {pages} {134113} (\bibinfo {year} {2009})},\ \Eprint
  {https://arxiv.org/abs/https://doi.org/10.1063/1.3103930}
  {https://doi.org/10.1063/1.3103930} \BibitemShut {NoStop}%
\bibitem [{\citenamefont {Symonds}, \citenamefont {Kattirtzi},\ and\
  \citenamefont {Shalashilin}(2018)}]{Symonds18}%
  \BibitemOpen
  \bibfield  {author} {\bibinfo {author} {\bibfnamefont {C.}~\bibnamefont
  {Symonds}}, \bibinfo {author} {\bibfnamefont {J.~A.}\ \bibnamefont
  {Kattirtzi}},\ and\ \bibinfo {author} {\bibfnamefont {D.~V.}\ \bibnamefont
  {Shalashilin}},\ }\bibfield  {title} {\enquote {\bibinfo {title} {The effect
  of sampling techniques used in the multiconfigurational {E}hrenfest
  method},}\ }\href {https://doi.org/10.1063/1.5020567} {\bibfield  {journal}
  {\bibinfo  {journal} {The Journal of Chemical Physics}\ }\textbf {\bibinfo
  {volume} {148}},\ \bibinfo {pages} {184113} (\bibinfo {year} {2018})},\
  \Eprint {https://arxiv.org/abs/https://doi.org/10.1063/1.5020567}
  {https://doi.org/10.1063/1.5020567} \BibitemShut {NoStop}%
\bibitem [{\citenamefont {Mignolet}\ and\ \citenamefont
  {Curchod}(2018)}]{Mignolet18}%
  \BibitemOpen
  \bibfield  {author} {\bibinfo {author} {\bibfnamefont {B.}~\bibnamefont
  {Mignolet}}\ and\ \bibinfo {author} {\bibfnamefont {B.~F.~E.}\ \bibnamefont
  {Curchod}},\ }\bibfield  {title} {\enquote {\bibinfo {title} {A walk through
  the approximations of ab initio multiple spawning},}\ }\href
  {https://doi.org/10.1063/1.5022877} {\bibfield  {journal} {\bibinfo
  {journal} {The Journal of Chemical Physics}\ }\textbf {\bibinfo {volume}
  {148}},\ \bibinfo {pages} {134110} (\bibinfo {year} {2018})},\ \Eprint
  {https://arxiv.org/abs/https://doi.org/10.1063/1.5022877}
  {https://doi.org/10.1063/1.5022877} \BibitemShut {NoStop}%
\bibitem [{\citenamefont {White}, \citenamefont {Tretiak},\ and\ \citenamefont
  {Mozyrsky}(2016)}]{White16}%
  \BibitemOpen
  \bibfield  {author} {\bibinfo {author} {\bibfnamefont {A.}~\bibnamefont
  {White}}, \bibinfo {author} {\bibfnamefont {S.}~\bibnamefont {Tretiak}},\
  and\ \bibinfo {author} {\bibfnamefont {D.}~\bibnamefont {Mozyrsky}},\
  }\bibfield  {title} {\enquote {\bibinfo {title} {Coupled wave-packets for
  non-adiabatic molecular dynamics: a generalization of gaussian wave-packet
  dynamics to multiple potential energy surfaces},}\ }\href
  {https://doi.org/10.1039/C6SC01319H} {\bibfield  {journal} {\bibinfo
  {journal} {Chem. Sci.}\ }\textbf {\bibinfo {volume} {7}},\ \bibinfo {pages}
  {4905--4911} (\bibinfo {year} {2016})}\BibitemShut {NoStop}%
\bibitem [{\citenamefont {Baskov}, \citenamefont {White},\ and\ \citenamefont
  {Mozyrsky}(2019)}]{Baskov2019}%
  \BibitemOpen
  \bibfield  {author} {\bibinfo {author} {\bibfnamefont {R.}~\bibnamefont
  {Baskov}}, \bibinfo {author} {\bibfnamefont {A.~J.}\ \bibnamefont {White}},\
  and\ \bibinfo {author} {\bibfnamefont {D.}~\bibnamefont {Mozyrsky}},\
  }\bibfield  {title} {\enquote {\bibinfo {title} {Improved {E}hrenfest
  approach to model correlated electron-nuclear dynamics},}\ }\href
  {https://doi.org/10.1021/acs.jpclett.8b03061} {\bibfield  {journal} {\bibinfo
   {journal} {The Journal of Physical Chemistry Letters}\ }\textbf {\bibinfo
  {volume} {10}},\ \bibinfo {pages} {433--440} (\bibinfo {year}
  {2019})}\BibitemShut {NoStop}%
\bibitem [{\citenamefont {Joubert-Doriol}\ and\ \citenamefont
  {Izmaylov}(2018)}]{Izmaylov18}%
  \BibitemOpen
  \bibfield  {author} {\bibinfo {author} {\bibfnamefont {L.}~\bibnamefont
  {Joubert-Doriol}}\ and\ \bibinfo {author} {\bibfnamefont {A.~F.}\
  \bibnamefont {Izmaylov}},\ }\bibfield  {title} {\enquote {\bibinfo {title}
  {Variational nonadiabatic dynamics in the moving crude adiabatic
  representation: Further merging of nuclear dynamics and electronic
  structure},}\ }\href {https://doi.org/10.1063/1.5020655} {\bibfield
  {journal} {\bibinfo  {journal} {The Journal of Chemical Physics}\ }\textbf
  {\bibinfo {volume} {148}},\ \bibinfo {pages} {114102} (\bibinfo {year}
  {2018})},\ \Eprint {https://arxiv.org/abs/https://doi.org/10.1063/1.5020655}
  {https://doi.org/10.1063/1.5020655} \BibitemShut {NoStop}%
\bibitem [{\citenamefont {Mandal}, \citenamefont {Yamijala},\ and\
  \citenamefont {Huo}(2018)}]{Pengfei18}%
  \BibitemOpen
  \bibfield  {author} {\bibinfo {author} {\bibfnamefont {A.}~\bibnamefont
  {Mandal}}, \bibinfo {author} {\bibfnamefont {S.~S.}\ \bibnamefont
  {Yamijala}},\ and\ \bibinfo {author} {\bibfnamefont {P.}~\bibnamefont
  {Huo}},\ }\bibfield  {title} {\enquote {\bibinfo {title} {Quasi-diabatic
  representation for nonadiabatic dynamics propagation},}\ }\href
  {https://doi.org/10.1021/acs.jctc.7b01178} {\bibfield  {journal} {\bibinfo
  {journal} {Journal of Chemical Theory and Computation}\ }\textbf {\bibinfo
  {volume} {14}},\ \bibinfo {pages} {1828--1840} (\bibinfo {year} {2018})},\
  \bibinfo {note} {pMID: 29489359},\ \Eprint
  {https://arxiv.org/abs/https://doi.org/10.1021/acs.jctc.7b01178}
  {https://doi.org/10.1021/acs.jctc.7b01178} \BibitemShut {NoStop}%
\bibitem [{\citenamefont {Tully}(1990)}]{Tully1990}%
  \BibitemOpen
  \bibfield  {author} {\bibinfo {author} {\bibfnamefont {J.~C.}\ \bibnamefont
  {Tully}},\ }\bibfield  {title} {\enquote {\bibinfo {title} {Molecular
  dynamics with electronic transitions},}\ }\href
  {https://doi.org/10.1063/1.459170} {\bibfield  {journal} {\bibinfo  {journal}
  {The Journal of Chemical Physics}\ }\textbf {\bibinfo {volume} {93}},\
  \bibinfo {pages} {1061--1071} (\bibinfo {year} {1990})}\BibitemShut {NoStop}%
\bibitem [{\citenamefont {Kapral}(2006)}]{Kapral06}%
  \BibitemOpen
  \bibfield  {author} {\bibinfo {author} {\bibfnamefont {R.}~\bibnamefont
  {Kapral}},\ }\bibfield  {title} {\enquote {\bibinfo {title} {Progress in the
  theory of mixed quantum-classical dynamics},}\ }\href
  {https://doi.org/10.1146/annurev.physchem.57.032905.104702} {\bibfield
  {journal} {\bibinfo  {journal} {Annual Review of Physical Chemistry}\
  }\textbf {\bibinfo {volume} {57}},\ \bibinfo {pages} {129--157} (\bibinfo
  {year} {2006})},\ \bibinfo {note} {pMID: 16599807},\ \Eprint
  {https://arxiv.org/abs/https://doi.org/10.1146/annurev.physchem.57.032905.104702}
  {https://doi.org/10.1146/annurev.physchem.57.032905.104702} \BibitemShut
  {NoStop}%
\bibitem [{\citenamefont {Parandekar}\ and\ \citenamefont
  {Tully}(2006)}]{Tully06}%
  \BibitemOpen
  \bibfield  {author} {\bibinfo {author} {\bibfnamefont {P.~V.}\ \bibnamefont
  {Parandekar}}\ and\ \bibinfo {author} {\bibfnamefont {J.~C.}\ \bibnamefont
  {Tully}},\ }\bibfield  {title} {\enquote {\bibinfo {title} {Detailed balance
  in {E}hrenfest mixed quantum-classical dynamics},}\ }\href
  {https://doi.org/10.1021/ct050213k} {\bibfield  {journal} {\bibinfo
  {journal} {Journal of Chemical Theory and Computation}\ }\textbf {\bibinfo
  {volume} {2}},\ \bibinfo {pages} {229--235} (\bibinfo {year} {2006})},\
  \bibinfo {note} {pMID: 26626509},\ \Eprint
  {https://arxiv.org/abs/https://doi.org/10.1021/ct050213k}
  {https://doi.org/10.1021/ct050213k} \BibitemShut {NoStop}%
\bibitem [{\citenamefont {Daligault}\ and\ \citenamefont
  {Mozyrsky}(2018)}]{Mozyrsky18}%
  \BibitemOpen
  \bibfield  {author} {\bibinfo {author} {\bibfnamefont {J.}~\bibnamefont
  {Daligault}}\ and\ \bibinfo {author} {\bibfnamefont {D.}~\bibnamefont
  {Mozyrsky}},\ }\bibfield  {title} {\enquote {\bibinfo {title} {Nonadiabatic
  quantum molecular dynamics with detailed balance},}\ }\href
  {https://doi.org/10.1103/PhysRevB.98.205120} {\bibfield  {journal} {\bibinfo
  {journal} {Phys. Rev. B}\ }\textbf {\bibinfo {volume} {98}},\ \bibinfo
  {pages} {205120} (\bibinfo {year} {2018})}\BibitemShut {NoStop}%
\bibitem [{\citenamefont {Sifain}, \citenamefont {Wang},\ and\ \citenamefont
  {Prezhdo}(2016)}]{Sifain16}%
  \BibitemOpen
  \bibfield  {author} {\bibinfo {author} {\bibfnamefont {A.~E.}\ \bibnamefont
  {Sifain}}, \bibinfo {author} {\bibfnamefont {L.}~\bibnamefont {Wang}},\ and\
  \bibinfo {author} {\bibfnamefont {O.~V.}\ \bibnamefont {Prezhdo}},\
  }\bibfield  {title} {\enquote {\bibinfo {title} {Communication: Proper
  treatment of classically forbidden electronic transitions significantly
  improves detailed balance in surface hopping},}\ }\href
  {https://doi.org/10.1063/1.4953444} {\bibfield  {journal} {\bibinfo
  {journal} {The Journal of Chemical Physics}\ }\textbf {\bibinfo {volume}
  {144}},\ \bibinfo {pages} {211102} (\bibinfo {year} {2016})},\ \Eprint
  {https://arxiv.org/abs/https://doi.org/10.1063/1.4953444}
  {https://doi.org/10.1063/1.4953444} \BibitemShut {NoStop}%
\bibitem [{\citenamefont {Jain}\ and\ \citenamefont {Subotnik}(2015)}]{Jain15}%
  \BibitemOpen
  \bibfield  {author} {\bibinfo {author} {\bibfnamefont {A.}~\bibnamefont
  {Jain}}\ and\ \bibinfo {author} {\bibfnamefont {J.~E.}\ \bibnamefont
  {Subotnik}},\ }\bibfield  {title} {\enquote {\bibinfo {title} {Surface
  hopping, transition state theory, and decoherence. ii. thermal rate constants
  and detailed balance},}\ }\href {https://doi.org/10.1063/1.4930549}
  {\bibfield  {journal} {\bibinfo  {journal} {The Journal of Chemical Physics}\
  }\textbf {\bibinfo {volume} {143}},\ \bibinfo {pages} {134107} (\bibinfo
  {year} {2015})},\ \Eprint
  {https://arxiv.org/abs/https://aip.scitation.org/doi/pdf/10.1063/1.4930549}
  {https://aip.scitation.org/doi/pdf/10.1063/1.4930549} \BibitemShut {NoStop}%
\bibitem [{\citenamefont {Kang}\ and\ \citenamefont {Wang}(2019)}]{Kang19}%
  \BibitemOpen
  \bibfield  {author} {\bibinfo {author} {\bibfnamefont {J.}~\bibnamefont
  {Kang}}\ and\ \bibinfo {author} {\bibfnamefont {L.-W.}\ \bibnamefont
  {Wang}},\ }\bibfield  {title} {\enquote {\bibinfo {title} {Nonadiabatic
  molecular dynamics with decoherence and detailed balance under a density
  matrix ensemble formalism},}\ }\href
  {https://doi.org/10.1103/PhysRevB.99.224303} {\bibfield  {journal} {\bibinfo
  {journal} {Phys. Rev. B}\ }\textbf {\bibinfo {volume} {99}},\ \bibinfo
  {pages} {224303} (\bibinfo {year} {2019})}\BibitemShut {NoStop}%
\bibitem [{\citenamefont {Miller}\ and\ \citenamefont
  {Cotton}(2015)}]{Miller15}%
  \BibitemOpen
  \bibfield  {author} {\bibinfo {author} {\bibfnamefont {W.~H.}\ \bibnamefont
  {Miller}}\ and\ \bibinfo {author} {\bibfnamefont {S.~J.}\ \bibnamefont
  {Cotton}},\ }\bibfield  {title} {\enquote {\bibinfo {title} {Communication:
  Note on detailed balance in symmetrical quasi-classical models for
  electronically non-adiabatic dynamics},}\ }\href
  {https://doi.org/10.1063/1.4916945} {\bibfield  {journal} {\bibinfo
  {journal} {The Journal of Chemical Physics}\ }\textbf {\bibinfo {volume}
  {142}},\ \bibinfo {pages} {131103} (\bibinfo {year} {2015})},\ \Eprint
  {https://arxiv.org/abs/https://doi.org/10.1063/1.4916945}
  {https://doi.org/10.1063/1.4916945} \BibitemShut {NoStop}%
\bibitem [{\citenamefont {Baer}(2006)}]{Baer2006}%
  \BibitemOpen
  \bibfield  {author} {\bibinfo {author} {\bibfnamefont {M.}~\bibnamefont
  {Baer}},\ }\href@noop {} {\emph {\bibinfo {title} {Beyond Born-Oppenheimer:
  Electronic Nonadiabatic Coupling Terms and Conical Intersections}}}\
  (\bibinfo  {publisher} {Wiley-Interscience},\ \bibinfo {year}
  {2006})\BibitemShut {NoStop}%
\bibitem [{\citenamefont {Levine}(2014)}]{Levine2014}%
  \BibitemOpen
  \bibfield  {author} {\bibinfo {author} {\bibfnamefont {I.~N.}\ \bibnamefont
  {Levine}},\ }\href@noop {} {\emph {\bibinfo {title} {Quantum Chemistry}}},\
  \bibinfo {edition} {7th}\ ed.\ (\bibinfo  {publisher} {Pearson},\ \bibinfo
  {year} {2014})\BibitemShut {NoStop}%
\bibitem [{\citenamefont {Heller}(1975)}]{Heller1975}%
  \BibitemOpen
  \bibfield  {author} {\bibinfo {author} {\bibfnamefont {E.~J.}\ \bibnamefont
  {Heller}},\ }\bibfield  {title} {\enquote {\bibinfo {title} {Time-dependent
  approach to semiclassical dynamics},}\ }\href
  {https://doi.org/10.1063/1.430620} {\bibfield  {journal} {\bibinfo  {journal}
  {The Journal of Chemical Physics}\ }\textbf {\bibinfo {volume} {62}},\
  \bibinfo {pages} {1544--1555} (\bibinfo {year} {1975})}\BibitemShut {NoStop}%
\bibitem [{\citenamefont {Landau}\ and\ \citenamefont
  {Lifshits}(1958)}]{Landau}%
  \BibitemOpen
  \bibfield  {author} {\bibinfo {author} {\bibfnamefont {L.~D.}\ \bibnamefont
  {Landau}}\ and\ \bibinfo {author} {\bibfnamefont {E.~M.}\ \bibnamefont
  {Lifshits}},\ }\href@noop {} {\emph {\bibinfo {title} {Quantum Mechanics}}},\
  Vol.~\bibinfo {volume} {3}\ (\bibinfo  {publisher} {Pergamon Press Inc.},\
  \bibinfo {year} {1958})\BibitemShut {NoStop}%
\bibitem [{\citenamefont {Stueckelberg}(1932)}]{Stueckelberg1932}%
  \BibitemOpen
  \bibfield  {author} {\bibinfo {author} {\bibfnamefont {E.~C.~G.}\
  \bibnamefont {Stueckelberg}},\ }\bibfield  {title} {\enquote {\bibinfo
  {title} {Theory of inelastic collisions between atoms},}\ }\href@noop {}
  {\bibfield  {journal} {\bibinfo  {journal} {Helvetica Physica Acta}\ }\textbf
  {\bibinfo {volume} {5}},\ \bibinfo {pages} {369--423} (\bibinfo {year}
  {1932})}\BibitemShut {NoStop}%
\bibitem [{\citenamefont {Schwartz}\ \emph {et~al.}(1996)\citenamefont
  {Schwartz}, \citenamefont {Bittner}, \citenamefont {Prezhdo},\ and\
  \citenamefont {Rossky}}]{Schwartz1996}%
  \BibitemOpen
  \bibfield  {author} {\bibinfo {author} {\bibfnamefont {B.~J.}\ \bibnamefont
  {Schwartz}}, \bibinfo {author} {\bibfnamefont {E.~R.}\ \bibnamefont
  {Bittner}}, \bibinfo {author} {\bibfnamefont {O.~V.}\ \bibnamefont
  {Prezhdo}},\ and\ \bibinfo {author} {\bibfnamefont {P.~J.}\ \bibnamefont
  {Rossky}},\ }\bibfield  {title} {\enquote {\bibinfo {title} {Quantum
  decoherence and the isotope effect in condensed phase nonadiabatic molecular
  dynamics simulations},}\ }\href {https://doi.org/10.1063/1.471326} {\bibfield
   {journal} {\bibinfo  {journal} {The Journal of Chemical Physics}\ }\textbf
  {\bibinfo {volume} {104}},\ \bibinfo {pages} {5942--5955} (\bibinfo {year}
  {1996})}\BibitemShut {NoStop}%
\bibitem [{\citenamefont {Jaeger}, \citenamefont {Fischer},\ and\ \citenamefont
  {Prezhdo}(2012)}]{Jaeger2012}%
  \BibitemOpen
  \bibfield  {author} {\bibinfo {author} {\bibfnamefont {H.~M.}\ \bibnamefont
  {Jaeger}}, \bibinfo {author} {\bibfnamefont {S.}~\bibnamefont {Fischer}},\
  and\ \bibinfo {author} {\bibfnamefont {O.~V.}\ \bibnamefont {Prezhdo}},\
  }\bibfield  {title} {\enquote {\bibinfo {title} {Decoherence-induced surface
  hopping},}\ }\href {https://doi.org/10.1063/1.4757100} {\bibfield  {journal}
  {\bibinfo  {journal} {The Journal of Chemical Physics}\ }\textbf {\bibinfo
  {volume} {137}},\ \bibinfo {pages} {22A545} (\bibinfo {year}
  {2012})}\BibitemShut {NoStop}%
\bibitem [{\citenamefont {Subotnik}(2010)}]{Subotnik2010}%
  \BibitemOpen
  \bibfield  {author} {\bibinfo {author} {\bibfnamefont {J.~E.}\ \bibnamefont
  {Subotnik}},\ }\bibfield  {title} {\enquote {\bibinfo {title} {Augmented
  {E}hrenfest dynamics yields a rate for surface hopping},}\ }\href
  {https://doi.org/10.1063/1.3314248} {\bibfield  {journal} {\bibinfo
  {journal} {The Journal of Chemical Physics}\ }\textbf {\bibinfo {volume}
  {132}},\ \bibinfo {pages} {134112} (\bibinfo {year} {2010})}\BibitemShut
  {NoStop}%
\bibitem [{\citenamefont {Subotnik}\ and\ \citenamefont
  {Shenvi}(2011)}]{Subotnik2011}%
  \BibitemOpen
  \bibfield  {author} {\bibinfo {author} {\bibfnamefont {J.~E.}\ \bibnamefont
  {Subotnik}}\ and\ \bibinfo {author} {\bibfnamefont {N.}~\bibnamefont
  {Shenvi}},\ }\bibfield  {title} {\enquote {\bibinfo {title} {A new approach
  to decoherence and momentum rescaling in the surface hopping algorithm},}\
  }\href {https://doi.org/10.1063/1.3506779} {\bibfield  {journal} {\bibinfo
  {journal} {The Journal of Chemical Physics}\ }\textbf {\bibinfo {volume}
  {134}},\ \bibinfo {pages} {024105} (\bibinfo {year} {2011})}\BibitemShut
  {NoStop}%
\bibitem [{\citenamefont {Shenvi}, \citenamefont {Subotnik},\ and\
  \citenamefont {Yang}(2011)}]{Shenvi2011}%
  \BibitemOpen
  \bibfield  {author} {\bibinfo {author} {\bibfnamefont {N.}~\bibnamefont
  {Shenvi}}, \bibinfo {author} {\bibfnamefont {J.~E.}\ \bibnamefont
  {Subotnik}},\ and\ \bibinfo {author} {\bibfnamefont {W.}~\bibnamefont
  {Yang}},\ }\bibfield  {title} {\enquote {\bibinfo {title}
  {Simultaneous-trajectory surface hopping: A parameter-free algorithm for
  implementing decoherence in nonadiabatic dynamics},}\ }\href
  {https://doi.org/10.1063/1.3575588} {\bibfield  {journal} {\bibinfo
  {journal} {The Journal of Chemical Physics}\ }\textbf {\bibinfo {volume}
  {134}},\ \bibinfo {pages} {144102} (\bibinfo {year} {2011})}\BibitemShut
  {NoStop}%
\bibitem [{\citenamefont {Shenvi}\ and\ \citenamefont
  {Yang}(2012)}]{Shenvi2012}%
  \BibitemOpen
  \bibfield  {author} {\bibinfo {author} {\bibfnamefont {N.}~\bibnamefont
  {Shenvi}}\ and\ \bibinfo {author} {\bibfnamefont {W.}~\bibnamefont {Yang}},\
  }\bibfield  {title} {\enquote {\bibinfo {title} {Achieving partial
  decoherence in surface hopping through phase correction},}\ }\href
  {https://doi.org/10.1063/1.4746407} {\bibfield  {journal} {\bibinfo
  {journal} {The Journal of Chemical Physics}\ }\textbf {\bibinfo {volume}
  {137}},\ \bibinfo {pages} {22A528} (\bibinfo {year} {2012})}\BibitemShut
  {NoStop}%
\bibitem [{\citenamefont {Subotnik}, \citenamefont {Ouyang},\ and\
  \citenamefont {Landry}(2013)}]{Subotnik2013}%
  \BibitemOpen
  \bibfield  {author} {\bibinfo {author} {\bibfnamefont {J.~E.}\ \bibnamefont
  {Subotnik}}, \bibinfo {author} {\bibfnamefont {W.}~\bibnamefont {Ouyang}},\
  and\ \bibinfo {author} {\bibfnamefont {B.~R.}\ \bibnamefont {Landry}},\
  }\bibfield  {title} {\enquote {\bibinfo {title} {Can we derive {T}ully's
  surface-hopping algorithm from the semiclassical quantum {L}iouville
  equation? almost, but only with decoherence},}\ }\href
  {https://doi.org/10.1063/1.4829856} {\bibfield  {journal} {\bibinfo
  {journal} {The Journal of Chemical Physics}\ }\textbf {\bibinfo {volume}
  {139}},\ \bibinfo {pages} {214107} (\bibinfo {year} {2013})}\BibitemShut
  {NoStop}%
\bibitem [{\citenamefont {Akimov}, \citenamefont {Long},\ and\ \citenamefont
  {Prezhdo}(2014)}]{Akimov2014}%
  \BibitemOpen
  \bibfield  {author} {\bibinfo {author} {\bibfnamefont {A.~V.}\ \bibnamefont
  {Akimov}}, \bibinfo {author} {\bibfnamefont {R.}~\bibnamefont {Long}},\ and\
  \bibinfo {author} {\bibfnamefont {O.~V.}\ \bibnamefont {Prezhdo}},\
  }\bibfield  {title} {\enquote {\bibinfo {title} {Coherence penalty
  functional: A simple method for adding decoherence in {E}hrenfest
  dynamics},}\ }\href {https://doi.org/10.1063/1.4875702} {\bibfield  {journal}
  {\bibinfo  {journal} {The Journal of Chemical Physics}\ }\textbf {\bibinfo
  {volume} {140}},\ \bibinfo {pages} {194107} (\bibinfo {year}
  {2014})}\BibitemShut {NoStop}%
\bibitem [{\citenamefont {Baer}(2002)}]{Baer2002}%
  \BibitemOpen
  \bibfield  {author} {\bibinfo {author} {\bibfnamefont {M.}~\bibnamefont
  {Baer}},\ }\bibfield  {title} {\enquote {\bibinfo {title} {Introduction to
  the theory of electronic non-adiabatic coupling terms in molecular
  systems},}\ }\href
  {https://doi.org/https://doi.org/10.1016/S0370-1573(01)00052-7} {\bibfield
  {journal} {\bibinfo  {journal} {Physics Reports}\ }\textbf {\bibinfo {volume}
  {358}},\ \bibinfo {pages} {75 -- 142} (\bibinfo {year} {2002})}\BibitemShut
  {NoStop}%
\bibitem [{\citenamefont {Gherib}\ \emph {et~al.}(2016)\citenamefont {Gherib},
  \citenamefont {Ye}, \citenamefont {Ryabinkin},\ and\ \citenamefont
  {Izmaylov}}]{Gherib2016}%
  \BibitemOpen
  \bibfield  {author} {\bibinfo {author} {\bibfnamefont {R.}~\bibnamefont
  {Gherib}}, \bibinfo {author} {\bibfnamefont {L.}~\bibnamefont {Ye}}, \bibinfo
  {author} {\bibfnamefont {I.~G.}\ \bibnamefont {Ryabinkin}},\ and\ \bibinfo
  {author} {\bibfnamefont {A.~F.}\ \bibnamefont {Izmaylov}},\ }\bibfield
  {title} {\enquote {\bibinfo {title} {On the inclusion of the diagonal
  {B}orn-{O}ppenheimer correction in surface hopping methods},}\ }\href
  {https://doi.org/10.1063/1.4945817} {\bibfield  {journal} {\bibinfo
  {journal} {The Journal of Chemical Physics}\ }\textbf {\bibinfo {volume}
  {144}},\ \bibinfo {pages} {154103} (\bibinfo {year} {2016})}\BibitemShut
  {NoStop}%
\bibitem [{\citenamefont {Jasper}\ \emph {et~al.}(2004)\citenamefont {Jasper},
  \citenamefont {Kendrick}, \citenamefont {Mead},\ and\ \citenamefont
  {Truhlar}}]{Jasper2004}%
  \BibitemOpen
  \bibfield  {author} {\bibinfo {author} {\bibfnamefont {A.}~\bibnamefont
  {Jasper}}, \bibinfo {author} {\bibfnamefont {B.~K.}\ \bibnamefont
  {Kendrick}}, \bibinfo {author} {\bibfnamefont {C.~A.}\ \bibnamefont {Mead}},\
  and\ \bibinfo {author} {\bibfnamefont {D.~G.}\ \bibnamefont {Truhlar}},\
  }\bibfield  {title} {\enquote {\bibinfo {title} {Non-{B}orn-{O}ppenheimer
  chemistry: Potential surfaces, couplings, and dynamics},}\ }in\ \href
  {https://doi.org/10.1142/9789812565426_0008} {\emph {\bibinfo {booktitle}
  {Modern Trends in Chemical Reaction Dynamics Part I}}},\ \bibinfo {editor}
  {edited by\ \bibinfo {editor} {\bibfnamefont {X.}~\bibnamefont {Yang}}\ and\
  \bibinfo {editor} {\bibfnamefont {L.}~\bibnamefont {K.}}}\ (\bibinfo
  {publisher} {World Scientific},\ \bibinfo {address} {Singapore},\ \bibinfo
  {year} {2004})\ Chap.~\bibinfo {chapter} {8}, pp.\ \bibinfo {pages}
  {329--391}\BibitemShut {NoStop}%
\bibitem [{\citenamefont {Takatsuka}(2007)}]{Takatsuka2007}%
  \BibitemOpen
  \bibfield  {author} {\bibinfo {author} {\bibfnamefont {K.}~\bibnamefont
  {Takatsuka}},\ }\bibfield  {title} {\enquote {\bibinfo {title}
  {Generalization of classical mechanics for nuclear motions on
  nonadiabatically coupled potential energy surfaces in chemical reactions},}\
  }\href {https://doi.org/10.1021/jp072233j} {\bibfield  {journal} {\bibinfo
  {journal} {The Journal of Physical Chemistry A}\ }\textbf {\bibinfo {volume}
  {111}},\ \bibinfo {pages} {10196--10204} (\bibinfo {year}
  {2007})}\BibitemShut {NoStop}%
\bibitem [{\citenamefont {U'Ren}, \citenamefont {Banaszek},\ and\ \citenamefont
  {Walmsley}(2003)}]{URen2003}%
  \BibitemOpen
  \bibfield  {author} {\bibinfo {author} {\bibfnamefont {A.~B.}\ \bibnamefont
  {U'Ren}}, \bibinfo {author} {\bibfnamefont {K.}~\bibnamefont {Banaszek}},\
  and\ \bibinfo {author} {\bibfnamefont {I.~A.}\ \bibnamefont {Walmsley}},\
  }\bibfield  {title} {\enquote {\bibinfo {title} {Photon engineering for
  quantum information processing},}\ }\href@noop {} {\bibfield  {journal}
  {\bibinfo  {journal} {Quantum Inf. Comput.}\ }\textbf {\bibinfo {volume}
  {3}},\ \bibinfo {pages} {480--502} (\bibinfo {year} {2003})},\ \Eprint
  {https://arxiv.org/abs/quant-ph/0305192} {quant-ph/0305192} \BibitemShut
  {NoStop}%
\end{thebibliography}%
\end{document}